\documentclass[reqno,a4paper]{amsart}
\usepackage{amssymb,stmaryrd}
\usepackage{amsfonts}
\usepackage{amstext}
\usepackage{color}
\usepackage{graphicx}
\usepackage{subcaption}
\usepackage{hyperref}
\usepackage{amsmath}

\newtheorem{theorem}{Theorem}[section]

\newtheorem{problem}[theorem]{Problem}

\newcommand{\extd}{\mathrm{d}}
\newcommand{\id}{\mathrm{id}}

\newcommand{\tens}{\otimes}

\newcommand{\del}{\partial}

\newcommand{\C}{\mathbb{C}}
\newcommand{\N}{\mathbb{N}}
\newcommand{\R}{\mathbb{R}}
\newcommand{\Z}{\mathbb{Z}}

\newcommand{\CJ}{\mathcal{J}}

\newcommand{\CF}{\mathcal{F}}
\newcommand{\CO}{\mathcal{O}}
\newcommand{\CD}{\mathcal{D}}

\newcommand{\cg}{\mathfrak{g}}
\newcommand{\ch}{\mathfrak{h}}

\newcommand{\eps}{\epsilon}
\newcommand{\<}{\langle}
\renewcommand{\>}{\rangle}

\newcommand{\cross}{{>\!\!\!\triangleleft}}

\newcommand{\bicross}{{\blacktriangleright\!\!\!\triangleleft}}

\renewcommand{\imath}{{\mathfrak{i}}}
\renewcommand{\div}{{\mathrm{div}}}

\begin{document}

\title{Algebraic approach to quantum gravity IV: applications}

\author[Shahn Majid]{\large Shahn Majid \\ \ \\ 
  School of Mathematical Sciences\\ 
  Queen Mary University of London\\ 
  Mile End Road, London E1 4NS, UK}

\thanks{
{\it Authors to whom correspondence should be addressed:}  s.majid@qmul.ac.uk}

\thanks{{\it Funding:} work supported by a Leverhulme Trust project grant RPG-2024-177}
\date{Mar 2026}



\date{\today}

\begingroup
\let\MakeUppercase\relax 
\maketitle
\endgroup

\author{Shahn Majid}

\begin{abstract}We provide a relatively self-contained introduction to the application of quantum spacetime and quantum Riemannian geometry to theoretical physics. Recent successes include calculation of the vacuum energy of spacetime curvature fluctuations in a single-plaquette model of quantum gravity, derivation of the Kaluza-Klein ansatz as a consequence of quantum spacetime, exactly conserved Noether charges from variational calculus on a lattice, and a new theory of classical and quantum geodesics. The latter leads to a theory of generally covariant quantum mechanics applicable in General Relativity with intriguing first results for the case of a black-hole.  We discuss several open problems past and present, and how they might be addressed going forward. New results include a phase transition for Euclidean quantum gravity on a 4-pointed star. 
\end{abstract}


\section{Introduction}

Quantum groups\cite{Dri,Ma} and various forms of noncommutative geometry have been around now for more than 40 years but their impact in mainstream theoretical physics remains limited. Certain quantum integrable systems and TQFT's, including 2+1 quantum gravity indeed have quantum group symmetries but these are specific models rather than telling us about theoretical physics more generally. Similarly, the quantum-spacetime hypothesis that spacetime coordinates are plausibly better modelled as noncommutative (in the manner of quantum observables) due to quantum gravity effects\cite{Sny,Ma:pla, DFR,MaRue,Hoo} opened the door to a flood of concrete models,  but without a clear understanding of how to connect such algebra to the real world and hence to mainstream physics. As I will try to explain, this is now starting to change. My own thinking on quantum gravity here is charted in previous surveys\cite{Ma:mea, Ma:qg3,Ma:non,Ma:qg1}, where the last of these is philosophical and the others are rather more concrete.

In these notes, I will be relatively light on technicalities and only recall the framework of noncommutative geometry that we use\cite{BegMa} in outline form, but with links for further reading. There will be new results in Section~\ref{sec2},  where I compute a baby Euclidean quantum gravity model on a 4-pointed star, in Section~\ref{sec3}, where I revisit quantum gravity on a square in more detail than elsewhere, and in Section~\ref{sec5}, where I look in more detail at entropy and now also relative entropy around a black-hole. For the most part, however, the focus will be on conceptual issues and how some long-standing ones might be solved going forward. In fact there are at least three approaches to noncommutative geometry: one coming out of operator algebras\cite{Con}, a more algebraic approach coming out quantum groups but not limited to them (which is the one we use, quantum Riemannian geometry or QRG), and noncommutative algebraic geometry motivated by sheaves and methods of algebraic geometry. A brief introduction to QRG is given below. Without going into details,  some tangible applications so far are:
\begin{itemize}\item Vacuum energy from curvature fluctuations on a square\cite{BliMa} and solution to the problem of the cosmological constant in the approach of  \cite{Car,WanUn}.
\item Noncommutative origin of the KK ansatz leading to gravity + Yang-Mills \cite{ArgMa3,LiuMa2,LiuMa3,LiuMa4,LiuMa5}.
\item Generally covariant quantum mechanics\cite{BegMa:qm, BegMa:flrw,KumMa} with gravatom states and new effects at the horizon in the case of a black-hole. 
\item New geodesic flow tools for ordinary GR and tests of the hypothesis of amplitudes on spacetime \cite{KumMa}.
\item Exactly conserved quantities for scalar lattice field theory from variational calculus on a lattice\cite{MaSim3}.
\end{itemize}

We will cover all of these in some form. As a teaser, let me note that even in the original Kaluza-Klein (KK) theory, landing on the right values for electromagnetism on spacetime needs an internal fibre circle of 23 Planck lengths. Something of this size ought to have significant quantum gravity corrections and it turns out that these actually lead to the KK ansatz and provide its origin.   Most sections of the notes will include further specific problems that could be looked at. We then look more generally in Section~\ref{sec6} to various issues and themes that could be attacked going forward. These include include issues that remain even for flat quantum spacetime models, and issues for quantum field theory on them and on lattices if we want to build on \cite{MaSim3}.

\subsection{Elements of QRG} 

A gentle intro to the QRG formalism of \cite{BegMa} is in my previous Corfu proceedings\cite{ArgMa3}, hence this section will be only a short recap. On the other hand, one can skip this section and refer back to it only where needed. 

Our middle ground for noncommutative geometry has as starting point any possibly noncommutative `coordinate algebra' $A$ in the role of $C^\infty(M)$ where $M$ classically could be a smooth manifold, but now without necessarily worrying about completions. If $M$ has global coordinates $x^\mu$ then we could let these non-commute and take $A$ polynomials in these generators modulo commutation relations, but for physics we will generally want a bigger algebra that also includes noncommutative versions of other functions such as exponentials for place waves. The second main ingredient is a differential structure expressed algebraically as a vector space $\Omega^1$ of `1-forms'. If there are global coordinates then $\Omega^1$ could be generated over $A$ by $\extd x^\mu$,  but now with some commutation relations. Everything should ideally stay associative so $a(\omega b)=(a\omega)b$ for all $a,b\in A$ and $\omega\in \Omega^1$ (one says that $\Omega^1$ is an $A$-bimodule). There is a map $\extd:A
\to \Omega^1$ obeying the product rule $\extd(ab)=a(\extd b)+(\extd a)b$. The triple $(A,\Omega^1,\extd)$ replaces the role of a manifold at the crudest level of having a differential structure.

The next layer, needed for physics, is a metric. This is expressed as $\cg\in \Omega^1\tens_A\Omega^1$  (where $\tens_A$ means we identify $\omega a\tens \eta=\omega \tens a\eta$ for all $a\in A$) with an appropriate inverse metric $(\ ,\ ):\Omega^1\tens_A\Omega^1$ subject to some axioms\cite{BegMa:gra,BegMa}.  At this point, the effects of noncommutativity already start to make things wierd if we adopt the strongest version of the axioms (one can of course weaken them), namely that $(\ ,\ )$ descends to $\tens_A$ as stated and is a bimodule map, i.e. 
\begin{equation}\label{invmetric} a(\omega,\eta)=(a\omega,\eta),\quad (\omega a,\eta)=(\omega,a\eta),\quad (\omega,\eta a)=(\omega,\eta)a\end{equation}
for all $a\in A$, $\omega,\eta\in\Omega^1$. It then turns out that\cite{BegMa:gra} 
\begin{equation}\label{gcen} a \cg= \cg a\end{equation}
for all $a\in A$. When the algebra and differential calculus is very noncommutative, this can put a lot of constraints on $\cg$ amounting in the case where $A$ is a deformation of $C^\infty(M)$ to only certain classical metrics being limits of ones on $A$ (i.e. quantisable). This restriction is key to both the quantum gravity calculation\cite{Ma:squ,BliMa} and to the noncommutative origin of the KK ansatz\cite{LiuMa2,LiuMa3,LiuMa4,LiuMa5}.. There should also be some form of symmetry condition but there appear to be different ways to do this in different contexts (the original one was $\wedge(\cg)=0$ under the wedge product of 1-forms). 

After the metric, one can look for a quantum Levi-Civita connection (QLC) as a map $\nabla:\Omega^1\to \Omega^1\tens_A\Omega^1$. In this context, a right vector field $X:\Omega^1\to A$, i.e., that respects the right action in the sense $X(\omega a)=X(\omega)a$, can be applied to the first output to give $\nabla_X:\Omega^1\to \Omega^1$ in the role classically of covariant derivative along $X$. A QLC needs to be torsion free and metric compatible. The first requires us to extend $(A,\Omega^1,\extd)$ to an exterior algebra $(\Omega,\extd)$ with forms of all degrees, where $\Omega^0=A$ and $\extd^2=0$. With the product forms denoted $\wedge$,  torsion free is expressed as $\wedge\nabla=\extd: A\to \Omega^2$. Metric compatibility has a natural weak form (called `cotorsion free') as
\[(\extd\tens \id- \id\wedge\nabla)\cg=0,\]
(then a torsion and cotorsion free connection is called a weak one or a WQLC). This weaker condition in the classical case lands on a partial (skew-symmetrized version) of metric compatibility. For the full-strength version of metric compatibility,  the approach in \cite{BegMa:gra,BegMa} is to assume that there is a bimodule map $\sigma:\Omega^1\tens_A\Omega^1\to \Omega^1\tens_A\Omega^1$ called the `generalised braiding' (classically it would be the flip map swapping the tensor factors) such that
\[ \nabla(\omega a)=\sigma(\omega\tens\extd a)+ (\nabla\omega)a\]
for all $a\in A, \omega\in \Omega^1$. Any left connection is required to obey the left Leibniz rule $\nabla(a\omega)=\extd a\tens \omega+a\nabla\omega$ but if it also admits $\sigma$ such that the above right Leibniz rule holds then we call $\nabla$ a `bimodule connection' (here $\sigma$ is uniquely determined if it exists, so this is just a further property of some left connections). The idea goes back to \cite{DVM, Mou}. The concept applies similarly to bimodule connections on any bimodule. The nice thing about bimodule connections in general is that they are closed under tensor product, so in our case $\Omega^1\tens_A\Omega^1$ gets a bimodule connection and we can define metric compatibility as $\nabla(\cg)=0$ with respect to this.  

Finally, any left connection has a Riemann curvature $R_\nabla:\Omega^1\to \Omega^2\tens\Omega^1$. With a little more structure, such as a bimodule `lifting map' $i:\Omega^2\to \Omega^1\tens_A\Omega^2$ such that $\wedge\circ i=\id$ (classically this map just expresses a 2-form as an antisymmetric tensor), we can take a trace to define the Ricci tensor ${\rm Ricci}\in \Omega^1\tens_A\Omega^1$, and then apply $(\ ,\ )$ to get to the Ricci scalar. This allows us then, with appropriate integration measures on $A$ and on the moduli of QRGs, to write down quantum gravity on any algebra in a functional-integral form where we integrate over all $(\cg,\nabla)$. (In many cases $\nabla$, as classically, is uniquely determined by $\cg$.) One way to take the trace here is to use the metric and inverse metric 
\[ {\rm Ricci}=((\ ,\ )\tens\id)(\id\tens (i\tens\id) R_\nabla)\cg\]
for which we only need that $(\ ,\ )$ is a right module map that descends to $\tens_A$, i.e. could drop the first of (\ref{invmetric}). On the other hand, Ricci here is merely a copy of the classical approach without any deeper understanding (and hence might not be the final answer for noncommutative geometry). It is referred to in \cite{BegMa} as a `working definition' in the absence of a proper (but gradually emerging) theory of noncommutative variational calculus. Given ${\rm Ricci}$, we then define  the Ricci scalar by
\[ R=(\ , \ ){\rm Ricci}\in A.\]
While copying the usual definition of the Einstein tensor does sometimes give a reasonable answer, there is, however, no convincing definition that is typically conserved in the sense of zero divergence with respect to $((\ ,\ )\tens\id)\nabla$ (and not clear if that is the property we are looking for due to our limited understanding of noncommutative variational calculus). Also note that our definitions when applied classically to $C^\infty(M)$ give $-1/2$ times the classical Ricci tensor and scalar. 

Finally, we work over $\C$ but require a $*$-structure which classically would encode real geometry. Thus, $A$ is a $*$-algebra, $\Omega$ is a graded-$*$ algebra, $\cg$ is hermitian in a certain sense and $\nabla$ is $*$-preserving in a certain sense. The latter two classically reduced to real coefficients if we work with self-adjoint (i.e., what would classically be real) coordinates. Details are in \cite{BegMa}.

\section{Quantum gravity on the 4-pointed star and phase transition}\label{sec2}

In this section, we present new results computing Euclidean quantum gravity on a 4-pointed star, to add to the growing repertoire of solved models. We find remarkable parallels to the detailed study of quantum gravity on the fuzzy sphere in \cite{Ma:are}, including a phase transition.  Baby quantum gravity models such as this can be viewed either as indicative of the smallest scale structure of spacetime (which is the point of view in the recent application\cite{BliMa} covered in Section~\ref{sec3}), or in their own right as extremely simplified toy models of the Universe globally. Both points of view have their merits. For example, one could imagine discrete quantum gravity as a sum over all graphs and for each graph  an integral over all QRGs on it. At the extreme that approaches the continuum, you would have very large graphs, but at the other extreme you would have small graphs as a piece of the story. On the other hand, the problem of enumerating all graphs is an open problem in mathematics, and we would also have to invent a suitable measure, possibly taking inspiration from causal sets\cite{Dow}. 

\subsection{Viewing a discrete space as a QRG} 

Any graph is a QRG as far as the construction of a noncommutative geometry and metric are concerned\cite{Ma:gra} but without a general theory of existence or uniqueness of the QLC (which has to be solved for on a case by case basis). The idea is that if $A=\C(X)$ for a discrete set $X$ then all possible $(\Omega^1,\extd)$ on it are in 1-1 correspondence with graphs whose vertex set is $X$. The arrows literally provide a vector space basis $\{\omega_{x\to y}\}$ of $\Omega^1$ and the bimodule products and $\extd$ are given by:
\[  f.\omega_{x\to y}=f(x)\omega_{x\to y}, \quad \omega_{x\to y}.f=f(y)\omega_{x\to y},\quad \extd f=\sum_{x\to y} (f(y)-f(x))\omega_{x\to y}.\]
The $*$ operation on functions is complex conjugation of the value and on arrows is $\omega_{x\to y}^*=-\omega_{y\to x}$, which requires the graph to be bidirected (every arrow has a reverse arrow). In this case, the most general metric has the form\cite{Ma:gra}
\[ \cg=\sum_{x\to y} g_{x\to y} \omega_{x\to y}\tens\omega_{y\to x},\quad g_{x\to y}\in \R\setminus\{0\}.\]
The centrality of the metric (\ref{gcen}) requires the second arrow to end at the same $x$ as the start of the first arrow, and it is only due to this that a metric in QRG conforms to the intuition in discrete geometry that a metric should be a nonzero real weight on every arrow.  A natural `symmetry' condition in this context is that $g_{x\to y}=g_{y\to x}$ (so it depends only on the edge and not on the direction) but a curious feature of some graphs is that for a QLC to exist we might have to have a larger magnitude pointing into the bulk from a free end compared to pointing out\cite{ArgMa2, BegMa:gra}. Finally, to have a notion of torsion, we need to fix the higher $(\Omega,\extd)$ and a natural proposal for any graph is $\Omega_{min}$, where we impose quadratic relations 
\[ \sum_{z: x\to z\to y} \omega_{x\to z}\wedge \omega_{z\to y}\]
for all $x,y$ that are two steps apart. This is an inner calculus with $\extd=[\theta,\ \}$ where $\theta=\sum_{x\to y}\omega_{x\to y}$. If we put conditions on $x,y$ then we have bigger calculi, including a biggest `maximal prolongation' $\Omega_{max}$ (all others are quotients of it). Depending on the graph, $\Omega_{min}$ could have further quotients of interest. We are also allowed to have some edges positive and some negative. What in the literature is called `Euclidean' signature is all arrows positive. More precisely, on  inspection of the continuum limit in several models, one can see that this corresponds to a fully negative signature, i.e. one should $-\cg$ for the actual metric. This overall minus sign is usually  ignored as it does change the QLC if this exists. 

\begin{problem}\rm Formulate and study a notion of `Lorentzian' metric on a graph, for example with a stratification such that edges between slices are negative and along slices positive. We will see an example in the next section, but the concept should be developed more generally and, plausibly, related to causal sets\cite{Dow} on choosing a consistent preferred `positive' arrow direction for the negative edges.
\end{problem}

\subsection{QRG of the 4-pointed star} 

This  is a self-contained exercise in which we compute quantum gravity on the 4-pointed star. It can also be done for the 3-pointed star, and has been done for the 2-pointed star as this is the same as the 3-node chain in \cite{ArgMa2}. We number the vertices as 0 in the centre then 1-4 for the external nodes: 
\[  \begin{array}{ccccc}  &  & 1 & & \\ & & | & & \\ 4 & -\kern-5pt - & 0 & -\kern-5pt- & 2 \\ & & | & &\\ & & 3 & &\end{array}\]
The algebra $A$ is the algebra of functions in these five vertices. $\Omega^1$ is 8-dimensional as a vector space, being labelled by the arrows. For the exterior algebra, we use $\Omega_{min}$ which in our case leads to $\Omega^2$ being 3-dimensional as a vector space, namely given by four elements $\omega_i$ for $i=1,2,3,4$ and one relation, along with differential,
\[ \omega_i:=\omega_{0\to i}\wedge\omega_{i\to 0}, \quad\sum_{i=1}^4\omega_i=0,\quad \extd \omega_{0\to i}=\extd\omega_{i\to 0}=\omega_i.\]
Other products $\omega_{i\to 0}\wedge\omega_{0\to j}=0$, so these are the only 2-forms. This also means that $\Omega^3=0$, i.e. there are no 3-forms of higher. For a metric we have in principle 8 real weights but the existence of a QLC constrains them so that only $g_{i\to 0}$, say, are independent, with
\[ g_{0\to i}={g_{i\to 0}\over 2}\]
as the forced modified edge symmetry at the boundary. See \cite{BegMa:gra}, remembering that $\lambda_{x\to y}=1/g_{y\to x}$. We will take the Euclidean signature where $g_{i\to 0}>0$ (with $-\cg$ as the more physical metric where relevant). There is a unique QLC \cite[Thm. 3.4]{BegMa:gra}, setting $s=-1$ there:
\begin{align*} \nabla(\omega_{0\to i})&=\omega_{i\to 0}\tens\omega_{0\to i}+ {1\over 2}\omega_{0\to i}\tens\omega_{i\to 0}+\sum_{j\ne i}\left(\omega_{j\to 0}\tens\omega_{0\to i}-{1\over 2}\omega_{0\to j}\tens\omega_{j\to 0}\right)\\
 \nabla(\omega_{i\to 0})&=\omega_{i\to 0}\tens\omega_{0\to i}+ \omega_{0\to i}\tens\omega_{i\to 0}-\sum_{j\ne i}{g_{j\to 0}\over g_{i\to 0}}\omega_{i\to 0}\tens\omega_{0\to j}.
\end{align*}
We now `crank the handle' and compute the Riemann curvature, which comes out as 
\begin{align*} R_\nabla(\omega_{0\to i})&=-{\omega_i\over 2}\tens\left(\omega_{0\to i}-\sum_{j\ne i}{g_{j\to 0}\over g_{i\to 0}}\omega_{0\to j}\right)+\sum_{j\ne i}{\omega_j\over 2}\tens \left(\omega_{0\to j}-\sum_{k\ne j}{g_{k\to 0}\over g_{j\to 0}}\omega_{0\to k}\right)\\
R_\nabla(\omega_{i\to 0})&=0\end{align*}
after a tedious calculation. For the Ricci curvature, we need a lifting map $i:\Omega^2\to \Omega^1\tens_A\Omega^1$ and we take
\[ i(\omega_i)=\omega_{0\to i}\tens\omega_{i\to 0}-{1\over 4}\sum_{j=1}^4\omega_{0\to j}\tens\omega_{j\to 0}.\]
where the subtraction of the average is needed to respect the relation $\sum_i\omega_i=0$. After another straightforward calculation, we obtain
\begin{align*} {\rm Ricci}&=\omega_{i\to 0}\tens\omega_{0\to i}\left( g_{i\to 0}g^{-1}_{av}-1\right)+\omega_{i\to 0}\tens \omega_{0\to j}\sum_{j\ne i}\left( g_{j\to 0}g^{-1}_{av}+ {1\over 2}({g_{j\to 0}\over g_{i\to 0}}-1)\right) \\
R(i)& =-2 \left({1\over g_{i\to 0}}- g^{-1}_{av}\right),\quad R(0)=0;\quad  g^{-1}_{av}:={1\over 4}\sum_{k=1}^4 {1\over g_{k\to 0}},\end{align*}
where used the average inverse metric. 

Finally, for the Einstein-Hilbert action we need to choose a measure $\mu$ or weight when we sum over the vertices, built form the metric functions. In our 1-dimensional case a natural choice as in \cite{Ma:par} for the integer line is to just use the metric coefficient function, in our case $g_i:=g_{i\to 0}$ regarded as the value at vertex $i$ (and picking something, such as the average around these vertices, for the value at 0). Correcting also for the $1/2$ in the normalisation of the Ricci scalar, this gives the Einstein-Hilbert action as
\[ S_g=2\sum_{i}g_i R(i)=-12+ \sum_{i\ne j} {g_j\over g_i},\]
where we sum over the 12 cases where $i\ne j$.  It is worth noting that if we write $g_i=e^{\phi_i}$ for a real-valued `Liouville field' and expand for weak fields then we have
\[ S_g=-12+2\sum_{(i,j)} \cosh(\phi_i-\phi_j)=\sum_{(i,j)} (\phi_i-\phi_j)^2+\cdots, \]
where we sum over the six un-ordered paired. So, $S_g$ measures how much the metric values around the external legs differ from each other. In either case, we discard the -12 since this just changes the partition function by a constant factor. 

For quantum gravity from our functional integral perspective, we now need to choose a measure for the metric field integration. Based on recent experience in \cite{LiuMa5} in the context of quantum gravity on a fuzzy sphere, adapted to our case, we (i) directly vary the metric values as usual (this was the default for graph quantum gravity models so far), hence $\extd g_i$ for each $i=1,\cdots,4$; (ii)  use {\em Liouville measure}  $g_i^{-1}\extd g_i=\extd\phi_i$ for each $i=1,2,3,4$. We report the results for both, obtained numerically using Mathematica.  Integrations were done with PrecisionGoal=4, MaxRecursion=15 and, for Figure~\ref{star}(a), with WorkingPresicision=20. A transient spike in the $L\Delta R(i)$ plot there at $G=8.2$ was removed by hand (as an artefact removable with more precision). There is still a degree of numerical noise visible in the plots, which should be ignored given the four-fold iterated integration.

\subsection{Quantum gravity with direct measure} The quantum gravity theory has partition function
\[ Z=\int_\eps^L\extd g_1\cdots\extd g_4 e^{-{2\over G}\sum_{i\ne j}{g_j\over g_i}}=L^4\int_{\bar\eps}^1\extd \bar g_1\cdots\extd \bar g_4 e^{-{2\over G}\sum_{i\ne j}{\bar g_j\over \bar g_i}},\]
where $G$ is a dimensionless coupling constant (hence it is not exactly Newtons constant) and $L,\eps$ are IR and UV cutoffs in the metric values and hence have dimensions of area (length squared). The action is scale-invariant and we let $\bar g_i=g_i/L$ and $\bar\eps=\eps/L$ as the dimensionless quantities actually used in computations. Also in practice, $Z$ and also all expectations values of interest in this section appear to converge at $\eps=0$ on numerical integration, and hence we just set $\bar\eps=0$. Changing it to $10^{-20}$, say, does not change any of the plots at the level of accuracy used.  By definition (i.e. without having an actual operator picture) we define 
\[ \<\CO\>=Z^{-1}\int_\eps^L\extd g_1\cdots\extd g_4 e^{-{2\over G}\sum_{i\ne j}{g_j\over g_i}}\CO=L^4\int_{\bar\eps}^1\extd \bar g_1\cdots\extd \bar g_4 e^{-{2\over G}\sum_{i\ne j}{\bar g_j\over \bar g_i}}\CO,\]
where if a term in $\CO$ is homogenous of degree $m$ in the $g_i$ then in the second form we use the same expression as a function of $\bar g_i$ with an extra $L^m$ at the front. The plots are then shown with the scaling $L$ put back in.  Finally, although the $g_4$ integral (say) can be done analytically if we go to $L=\infty$ (one gets an Erf function) and two more integrals can be done numerically with apparent convergence going to $L=\infty$ for the quantities of interest, the final integral still needs $L$-regularisation. The plots are then quicker and less noisy but, on the other hand, such hybrid results break the symmetry between the $g_i$ variables, so we do not do this. 

\begin{figure}
\[\includegraphics[scale=0.65]{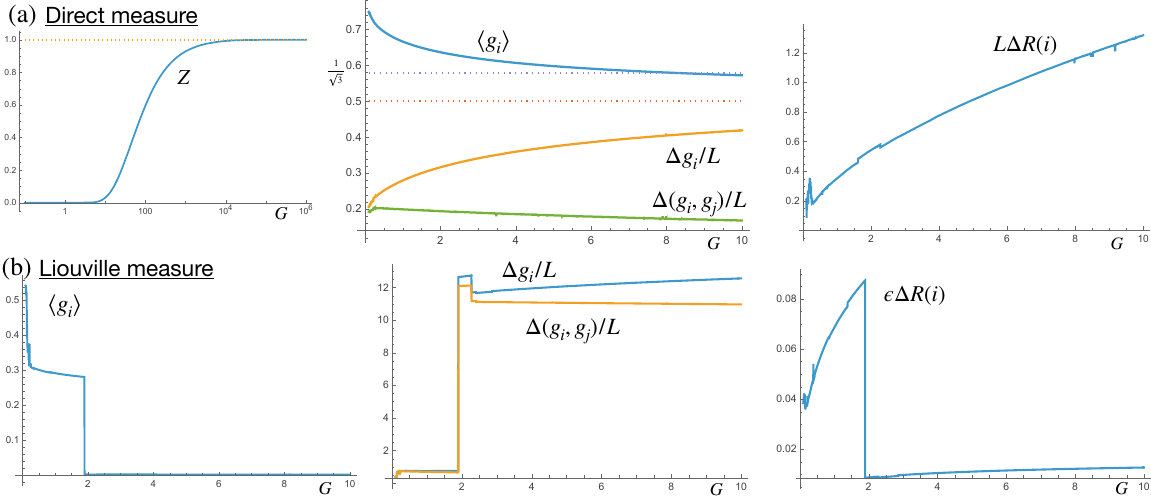}\]
\caption{Quantum gravity on a 4-pointed star graph for two different measures for the metric variables integration, as a function of the coupling constant $G$. The Liouville measure case (b) shows a phase transition at $G=2$. \label{star}}
\end{figure}

Results  are shown in Figure~\ref{star}(a) plotted against $G$. We see that $Z$ transitions from $0$ to $1$ (but note the log $G$-axis). Of interest in line with other models (see \cite{Ma:are}) is actually the relative uncertainty and relative mixed uncertainty (based on connected 2-point correlation functions), 
\[ \bar\Delta g_i={\Delta g_i\over \<g_i\>}=\sqrt{ {\<g_i^2\>-\<g_i\>^2\over \<g_i\>^2}},\quad \bar\Delta(g_i,g_j)=\sqrt{{\<g_i g_j\>-\<g_i\>\<g_j\>\over \<g_i\>\<g_j\>}}.\]
for $i\ne j$. We make use of the symmetry so its enough to calculate $\<g_1\>,\<g_1^2\>, \<g_1g_2\>$ for these. Note that for extremely large $G$, we can effectively ignore the action entirely and hence
\[ \<g_i\>\to {L\over 2},\quad \<g_i^2\>\to {L\over 3},\quad \<g_i g_j\>\to {L^2\over 4},\quad  \bar\Delta g_i\to {1\over \sqrt{3}},\quad \bar\Delta(g_i,g_j)\to 0\]
for $i\ne j$. These asymptotes are only hinted for the range of $G$ plotted (one has to go to $G=10^4$ to see them more clearly).  Symmetry of the expectation values also means that $\<R(i)\>=0$ while, looking at all the terms of $R(i)^2$ and again noting the symmetries when taking the expectation values, one can readily find that
\[\Delta R(i)={\sqrt{3}\over 2 }\sqrt{\<{1\over g_i^2}\> - \<{1\over g_i g_j}\>}\]
for $j\ne i$. This goes as $1/L$ when we work with dimensionless $\bar g_i$, and we  see from the plot that it increases gradually with $G$. A numerical fit for  large $G$  is 
\begin{equation}\label{DRsqG} \Delta R(i)\approx 0.542 {\sqrt{G}\over L}\end{equation}
as a good approximation to at least $G=10^5$. 

\subsection{Quantum gravity with Liouville measure} Now both the measure and the action are invariant under scaling, so we have that
\[ Z=\int_\eps^L{\extd g_1\cdots\extd g_4 \over g_1\cdots g_4} e^{-{2\over G}\sum_{i\ne j}{g_j\over g_i}}\]
depends only on the ratio $L/\eps$. Moreover, by a change of the four variables $g_i\to 1/g_i$, under which the action is also invariant, we find a remarkable duality
\[ \<\CO(g_1,\cdots,g_4)\>= \< \CO({L \eps\over g_1},\cdots, {L \eps\over g_4})\>.\]
Here, expectation values of a function $\CO$ in four variables as indicated are computed (on both sides) with $L,\eps$ as for $Z$ (and defined as usual by inserting the function in the integrand and dividing the result  by $Z$). We define $\Delta g_i$, $\Delta(g_i,g_j)$ as before but for $\Delta R(i)$ we now use the duality to compute instead
\begin{equation}\label{RDtot} \Delta R(i)={\sqrt{3}\over 2 L\eps}\sqrt{ \<g_i^2\>-\<g_i g_j\>}={\sqrt{3}\over 2 L\eps}\Delta_2 g_i,\end{equation}
where we define 
\[ \Delta_2 g_i:= \sqrt{(\Delta g_i)^2- (\Delta(g_i g_j))^2}\] 
for $i\ne j$ as the difference in the uncertainty attributable to the metric both from one variable and that from the connected correlations of two different metric variables.  In these terms, the original expression (which also applies for the direct measure) can be written as 
\begin{equation}\label{RDtotinv} \Delta R(i)={\sqrt{3}\over 2}\Delta_2({1\over g_i})\end{equation}
and in the Liouville measure case we use the duality to bring out the $L\eps$ before setting $\eps=0$ for the calculation of $\Delta_2 g_i$. After the transformation, all integrals of interest involve sufficiently positive powers of the $g_i$ so that we can set $\eps=0$ for all of the required expectation values. In fact, setting $\eps=0$ requires a caveat in the context of numerical integration. While Mathematica returns the integrals, it is in practice taking a slightly nonzero value when there are divergences involved, in our case (we shall see) around $\bar\eps=10^{-100}$. We will return to this theme in Section~\ref{sec5} that the internal precision for numerical computations is in fact similar to and a proxy for spacetime singularities being smoothed over at the Planck scale. As before, we do the calculations for $L=1$ with rescaled variables (and $\bar\eps=\eps/L$) and then scale to recover the remaining parameter $L$. 

The results, computed with $\eps=0$ (and with the above caveat) are shown in Figure~\ref{star}(b). We have not plotted $Z$ as it just increases for a log plot on the $G$-axis and can be (roughly) modelled for large $G$ as 
\[ Z\approx 5000 \sqrt{G}\]
up to machine limitations at around $G=10^4$. We also looked at the effect in $Z$ of setting an explicit value of $\eps$ and find say for $\bar\eps=10^{-10}$ that $Z$ drops by a factor of around 1/10 for larger $G$. We can model the correction factor as a function of $\bar\eps$ by
\[ Z(\bar\eps)\approx  Z  {\ln({1\over\bar\eps}^2)\over 241.2 \ln({G\over\bar\eps})},\]
as a decent fit (to within $\pm 5\%$) for at least $2<G<10^3$ and at least $\bar\eps>10^{-50}$. The formula is consistent with extrapolation to  $Z(10^{-110})$ being equated to $Z$ at $G=10$, as promised. This sensitivity to $\eps$ only applies for $G\ge 2$, plus  some small changes immediately prior to this value.  Moreover, none of the other integrals with positive powers of $g^i$ as observables are sensitive to $\eps$ if we explicitly introduce it. On the other hand, as  $Z$ occurs in the denominator, it means that the expectation values $\<g_i\>, \<g_ig_j\>$ scale by this factor, with knock-on effects to $\Delta(g_i), \Delta(g_i,g_j)$ and $\Delta R(i)$.  Thus, while the plots in Figure~\ref{star}(b) are nominally for $\eps=0$, we obtain qualitatively the same picture on explicitly introducing a finite value of $\eps$.

We see that  $\<g_i\>$ rapidly decreases and plateaus at about $0.3 L$ until $G=2$, then suddenly drops to a small value of around $0.003 L$, indicating a mild phase transition at $G=2$. At $\bar\eps=10^{-10}$, it looks the same but the drop is to about $0.03 L$. The values of  $\Delta g_i$ and $\Delta(g_i,g_j)$ for $i\ne j$ start at a similar value around $0.7 L$ before suddenly jumping at $G=2$ to around $12 L$  (with a smaller jump to $4.3 L$ for $\bar\eps=10^{-10}$).  
For $\Delta R(i)$, we similarly have $1/\eps$ times the plot shown, with a climb to $0.086/\eps$ and then a sudden drop at $G=2$ to around $0.009/\eps$ (or a smaller drop, to around $0.03/\eps$, for $\bar\eps=10^{-10}$). It then very gradually  increases, with a moderately good fit
\[ \eps\Delta R(i)\approx 0.0083 + 0.005 \ln(\ln(G))\]
(and $0.026 + 0.02 \ln(\ln(G))$ for $\bar\eps=10^{-10}$) up to $G=10^6$. 

The phase transition reported here is less extreme but similar to that for  Euclidean quantum gravity on the fuzzy sphere found in \cite{Ma:are} in that the jump depends on the explicit value of $\eps$ when this is introduced, being a larger transition for smaller values of it.  As for that model, its physical significance is less clear since the model is Euclidean, but merits further study.

\begin{problem}\rm Do a similar analysis for the 3-pointed star with QRG in \cite[Thm.~3.4]{BegMa:gra}. Also, while the QRG for the integer half-line $\N$ and the $n$-node chain $\bullet$-$\bullet$-$\cdots$-$\bullet$ with $\Omega_{min}$ is known\cite{ArgMa2} (the latter is a $q$-deformation of the former, for $q=e^{\imath\pi\over n+1}$), the resulting quantum gravity theory should be explored further for general $n$.  Also, the boundary effect that both models exhibit (that the metric cannot be edge symmetric but rather the metric value needs to be more pointing into the bulk) should be studied further in case there is a general result pertaining to a free `end' in a graph. \end{problem}

\section{Quantum gravity energy density of spacetime fluctuations}\label{sec3}

This section revisits quantum gravity on Lorentzian square\cite{Ma:squ} and recaps a recent application~\cite{BliMa} to the problem of the cosmological constant. The square here can be seen as a QRG in its own right in the spirit of Section~\ref{sec2}, but this time without any free ends (it is in some sense dual to the 4-pointed star). In the physical application in this section, however, we will think of it differently as indicative of the local structure at the Planck scale at a typical point in spacetime,  without worrying about exactly how it is repeated to assemble the larger picture. Ideally for this, we would like a hypercube to represent a cell in spacetime, but the calculations for that are formidable and have not yet been achieved. Also, this is not the same as a square lattice.

\subsection{Revisit of quantum gravity on a single Lorentzian plaqette}

The QRG for the Lorentzian square was first solved in \cite{Ma:squ} and we recap it with a certain amount of detail that previously  omitted. We then give more detailed plots of the metric field correlation functions for quantum gravity on the square, correcting an over-simplification in the original analysis (which is valid only in a certain limit), but with the same qualitative features.

As in \cite{Ma:squ}, we coordinatise the square written in short form with vertices $00,01,10,11$ arranged as 
 \[\begin{array}{ccc}
01 & -\kern-5pt-\kern-5pt-& 11\\ | & & | \\  00 & -\kern-5pt-\kern-5pt- &10\end{array} \]
We do not view this in $\R^2$ but as a discrete geometry in its own right. In fact,  we think of the vertices as $\Z_2\times\Z_2$ so that this is more like a discrete torus. Using this group structure, there is a 2-dimensional basis over $A=\C(\Z_2\times \Z_2)$ of left-invariant 1-forms, namely
\[ e^1=\omega_{00\to 10}+\omega_{10\to 00},\quad e^2=\omega_{00\to 01}+\omega_{01\to 00}\]
with non-commutation rules and a natural exterior algebra 
\[ e^i f=R_i(f)e^i,\quad \extd f=(\del_i f)e^i,\quad \del_i =R_i-\id,\quad  \{e^i,e^j\}=0,\quad \extd e^i=0,\]
for all $f\in A$, where 
\[  (R_1 f)(x,y)=f(x+1,y),\quad (R_2 f)(x,y)= f(x,y+1),\quad  xy\in 00,01,10,11\]
are the horizontal and vertical right translation operators when vertices are viewed in $\Z_2\times\Z_2$ (so addition here is mod 2). A quantum metric is $\cg\in \Omega^1\tens_A\Omega^1$ given by two functions $a,b\in A$
\[ \cg=-a e^1\tens e^1+ b e^2\tens e^2,\quad \del_1 a=\del_2 b=0,\]
where the conditions ensure that the `square lengths' are independent of the arrow direction. W take $a,b>0$ so that the square-lengths  $b_{00},b_{10}$ on the vertical edges are  positive, while $-a_{00}, -a_{01}$  on the horizontal edges are negative. The signature is technically $-+$ as opposed to $+-$, but this is ultimately a matter of conventions and we shall see  that the horizontal and vertical theories are anyway complex conjugates of each other.

In our case, the QLC has phase angle parameter $\theta$ according to $q=e^{\imath\theta}$ and is\cite{Ma:squ}
\[ \nabla e_1=(1+Q^{-1})e_1\tens e_1+(1-\alpha)(e_1\tens e_2+e_2\tens e_1)+ {b\over a}(R_2\beta-1)e_2\tens e_2,\]
\[ \nabla e_2={a\over b}(R_1\alpha-1)e_1\tens e_1+(1-\beta)(e_1\tens e_2+e_2\tens e_1)+(1-Q)e_2\tens e_2,\]
where $Q,\alpha,\beta$ are functions on the group defined as
\begin{equation}\label{Q} Q=(q,q^{-1},q^{-1},q),\quad \alpha=({a_{01}\over a_{00}}, 1, 1, {a_{00}\over a_{01}}),\quad \beta=(1, {b_{10}\over b_{00}},  {b_{00}\over b_{10}},1)\end{equation}
on the four vertices in the order stated above.  We write the Riemann curvature $R_\nabla(e^i)=\rho_{ij} e^1\wedge e^2\tens e^i$ as a 2-form valued operator on 1-forms with coefficients which can be computed as \cite{Ma:squ}
\begin{align*} \rho_{11}&=Q^{-1}R_1\alpha-Q\alpha+(1-\alpha)(R_1\beta-1)+{R_2 a\over a}(R_2\beta-1)(\alpha^{-1}-1),\\
\rho_{12}&=Q^{-1}(1-\alpha)+\alpha(R_2\alpha-1)-Q^{-1}{R_1b\over a}(\beta^{-1}-1)-{b\over a}(R_2\beta-1)R_2\beta,
\\ 
\rho_{21}&= Q(1-\beta)-\beta(R_1\beta-1)-Q{R_2a\over b}(\alpha^{-1}-1)+{a\over b}(R_1\alpha-1)R_1\alpha,  \\
\rho_{22}&= Q R_2\beta-Q^{-1}\beta+(\beta-1)(R_2\alpha-1)-{R_1 b\over b}(R_1\alpha-1)(\beta^{-1}-1),   \end{align*}
where the latter two are obtained  by a certain symmetry from the former two. The Ricci curvature depends on a lifting map $i:\Omega^2\to \Omega^1\tens_A\Omega^1$ and here there is an obvious one $i(e^1\wedge e^2)={1\over 2}(e^1\tens e^2-e^2\tens e^1)$, which  results in Ricci tensor and Ricci scalar
\[ {\rm Ricci}=R_{ij}e^i\tens e^j,\quad (R_{ij})={1\over 2}\begin{pmatrix}-R_2\rho_{21} & -R_2\rho_{22}\\ R_1\rho_{11} & R_1\rho_{12}\end{pmatrix},\]
\[ R={1\over 4 ab}\left(-(3+q+(1-q)\chi){\del^2 a\over\alpha}+(1-q^{-1}-(3+q^{-1})\chi) {\del^1 b\over\beta}\right).\]
Here,  $\chi(i,j)=(-1)^{i+j}$ and we corrected a sign typo in \cite{Ma:squ} in the 2nd expression. The Einstein-Hilbert action is\cite{Ma:squ}
\begin{equation}\label{EH} S_g=\sum_{\Z_2\times\Z_2} \mu R=(a_{00}-a_{01})^2({1\over a_{00}}+{1\over a_{01}})-(b_{00}-b_{10})^2({1\over b_{00}}+{1\over b_{10}})\end{equation}
which is independent of $q$ and computed with measure $\mu=ab>0$ as the natural choice from our data. If we had taken Euclidean signature with no minus sign for the 2nd term then this would have a  bathtub shape with minumum zero achieved at constant $a,b$.

Next, it is convenient to work with the $\Z_2\times\Z_2$ Fourier transformed metric values (the associated field momenta). This  amounts to a linear change of variables
\begin{equation}\label{fouab} a_{00}=k_0+k_1,\quad a_{01}=k_0-k_1,\quad b_{00}=l_0+l_1,\quad b_{10}=l_0-l_1\end{equation}
where $k_0,l_0>0$ are the average values of $a$ and $b$, and $|k_1|<k_0$ and $|l_1|< l_0$ so that we do not change the signature. It is also useful to work with the relative spatial field momenta 
\begin{equation}\label{kl} k={k_1\over k_0},\quad l={l_1\over l_0}\end{equation} 
where $|k|,|l|<1$. The Einstein-Hilbert action in these terms becomes\cite{Ma:squ} 
\begin{equation}\label{actionkl} S_g=k_0\alpha(k)- l_0\alpha(l);\quad \alpha(k):=  {8 k^2\over 1-k^2}\end{equation}
with square-length dimension, needing us to divide out by a coupling constant, which we denote $G$, also of square-length dimension.  Under this change of variables, the measure of integration becomes $\extd a_{00}\extd a_{01}\extd b_{00}\extd b_{10}=4\extd k_0\extd k_1\extd l_0\extd l_1=4  \extd k_0 \extd l_0 \extd k \extd l\, k_0 l_0$ and the partition function becomes $Z=|Z_1|^2$, where 
\begin{align*} 
Z_1&= 2\int_{-1}^1\extd k\int_\eps^L \extd k_0 k_0  e^{{\imath \over G} k_0 \alpha(k)}=- 4 G^2 \int_0^1\extd k{\extd\over\extd\alpha}|_{\alpha=\alpha(k)}{e^{ {\imath L\over G}\alpha}-e^{ {\imath \eps\over G}\alpha}\over \alpha}\\ &=- 16 G^2\int_0^\infty{\extd \alpha\over \alpha^{1\over 2}(8+\alpha)^{3\over 2}} {\extd\over\extd\alpha}\left({e^{ {\imath L\over G}\alpha}-e^{ {\imath \eps\over G}\alpha}\over \alpha}\right)\end{align*}
for the $k_0,k$ integration while the $l_0,l$ integration gives its complex conjugate. Here,  we regularised by limiting the $k_0$ integral to $\eps\le k_0\le L$, i.e. with both UV and IR cutoffs. In \cite{Ma:squ} we set $\eps=0$ as it did not affect the computations there, but we will need it for the curvature expectations in the next section so retain it.  We also note that $\alpha(k)$ is an even function and monotonic in the range $k\in [0,1)$, hence we changed variable to regard $k= \sqrt{\alpha\over 8+\alpha}$ as a function of $\alpha\in [0,\infty)$. For quantum gravity vacuum expectation values, we insert operators and by similar arguments as above, we have
\begin{equation}\label{kmf}  \<k_0^m f(\alpha(k))\>=(-\imath G)^m {\int_0^\infty {\extd \alpha\over \alpha^{1\over 2}(8+\alpha)^{3\over 2}}   f(\alpha) {\extd^{m+1}\over\extd\alpha^{m+1}}\left({e^{ {\imath L\over G}\alpha}-e^{ {\imath \eps\over G}\alpha}\over \alpha}\right)\over \int_0^\infty{\extd \alpha\over \alpha^{1\over 2}(8+\alpha)^{3\over 2}} {\extd\over\extd\alpha}\left({e^{ {\imath L\over G}\alpha}-e^{ {\imath \eps\over G}\alpha}\over \alpha}\right)},  \quad m\ge -1.\end{equation}
Inserting an odd function of $k$ in the original integral gives 0, after which it is sufficient to consider functions of $k^2$ and hence of $\alpha(k)$ as we have done. The complex conjugate results apply for the same function of $l_0,\alpha(l)$. This defines a baby quantum gravity theory on a single plaquette using a functional-integral approach. 

\begin{figure}
\[ \includegraphics[width=\textwidth]{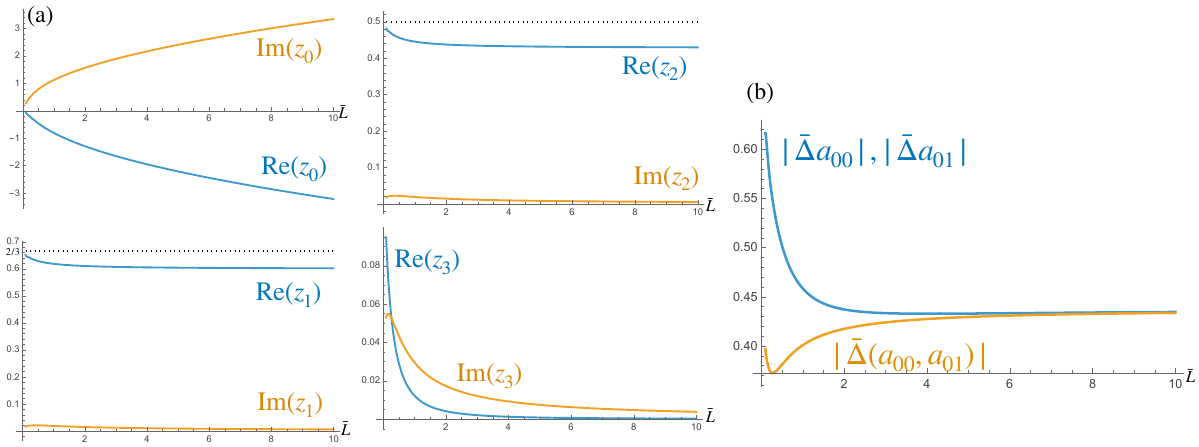}\]
\caption{(a) revisited partition and expectation value functions $z_i(\bar L)$ for quantum gravity on a square (b) relative uncertainty and correlators.\label{cor}}
\end{figure}

At this point, we should correct an error in \cite{Ma:squ} where it was stated that the integrals diverge at $\alpha=0$, which is not quite the case. On closer analysis, the numerical integration over $\alpha$ is highly unstable as we approach $\alpha=0$ and can appear misleading if the wrong machine precision is used, but we now observe that at least $Z_1$ can in fact be exactly solved. We let $\bar L=L/G$ as the relevant dimensionless quantity, then
\[Z_1=-\imath {G^2\over \sqrt{\pi}} \bar L\,  z_0(\bar L),\]
\[ z_0(\bar L)=(1-i) \sqrt{\bar L} \left(G_{2,3}^{2,2}\left(-8 i \bar L\left|
\begin{array}{c}
 \frac{1}{2},1 \\
 \frac{1}{2},\frac{3}{2},-\frac{1}{2} \\
\end{array}
\right.\right)+ G_{2,3}^{2,2}\left(-8 i \bar L\left|
\begin{array}{c}
 -\frac{1}{2},1 \\
 -\frac{1}{2},\frac{3}{2},-\frac{3}{2} \\
\end{array}
\right.\right)\right)+\sqrt{\pi }\]
in terms of MeijerG functions in the conventions of Mathematica. This is stated for $\eps=0$ but as a precaution we will in practice use $\bar\eps=\eps/G$ with value $10^{-5}$, in which case for the exact result we just subtract off the above with $\bar L$ replaced by $\bar\eps$. Its precise value does not affect any of the plots in this section. This exact result  (evaluated at WorkingPrecision=100 decimal places)  is then used as a reference point for the integrals that have to be done numerically and which are quite delicate. For these, we use WorkingPrecision=20 (or 30 when needed for $z_2, z_3$ below), and a sufficiently high value of "MaxErrorIncreases" in the "GlobalAdaptive" method. We also cutoff the integration over $\alpha$ at  $\delta=10^{-5}$ at the low end and at $1/\delta$ at the high end. This allows us to more reliably compute expectations as a function of $\bar L$ as
\[ \<a_{00}\>=\<a_{01}\>=\<k_0\>= L z_1(\bar L),\]
\[ \<a_{00}^2\>=\<a_{01}^2\>=\<k_0^2(1+k^2)\>= L^2 (z_2(\bar L)+z_3(\bar L)),\]
\[ \<a_{00}a_{01}\>=\<k_0^2(1-k^2)\>= L^2 (z_2(\bar L)-z_3(\bar L))\]
with the functions $z_i(\bar L)$ plotted in Figure~\ref{cor}. We see that for larger $\bar L$, 
\[z_1(\bar L)\approx 3/5, \quad z_2(\bar L)\approx  3/7,\quad  z_3(\bar L)\approx 0\]
not too far from the previous oversimplified analysis in \cite{Ma:squ}, where $z_1=2/3, z_2=1/2, z_3=0$ (and to which $z_1,z_2$ appear to approach for small $\bar L$.) Of interest are the {\em relative}  versions of the uncertainty of the metric and of the mixed uncertainties (measuring the connected correlators),
\[ \bar\Delta(a_{00})=\bar\Delta(a_{01})=\sqrt{\<a_{00}^2\>- \<a_{00}\>^2\over \<a_{00}\>^2}=\sqrt{{z_2(\bar L)+z_3(\bar L)\over z_1(\bar L)^2}-1},\]
\[ \bar\Delta(a_{00},a_{01})=\sqrt{\<a_{00}a_{01}\>-\<a_{00}\>\<a_{01}\>\over  \<a_{00}\>\<a_{01}\>}= \sqrt{{z_2(\bar L)-z_3(\bar L)\over z_1(\bar L)^2}-1}\]
as also plotted. Contrary to \cite{Ma:squ}, the expectation values are not real but they become real for large $\bar L$. There are similar results for the $l,l_0$ quantisation, related by complex conjugation. Our results now are more similar to other metric uncertainties and correlators in other models such as quantum gravity on $\Z_n$ and on the fuzzy sphere\cite{ArgMa,Ma:are}. Expectation values are regularised with respect to an upper limit $L$ of $k_0,l_0$, i.e. the maximum average value between the two parallel edges. In this sense, it is the maximum size of the plaquettes about which we further quantise the fluctuations. One can formally renormalise by fixing observed values such as $\<a_{00}\>$ at some $L$ and then writing in terms of this, but this does not really change much at larger $L$ since the two are more or less proportional. Hence we omit this, but see \cite{Ma:are,BliMa}.

\begin{problem}\rm Redo the quantum gravity theory for the same QRG  but with Liouville measure ${\extd a(00)\extd a(01)\over a(00)a(01)}$  as opposed to the direct measure used above. Also revisit both measures with the Euclidean square rather than the Lorentzian one above, i.e., with no $\imath$ in the action, to see if there is a phase transition. Questions of interpretation also remain unanswered even in the above baby model, in particular if there is a corresponding Hamiltonian quantisation or not (see discussion later). \end{problem}

\subsection{Calculation of vacuum energy of spacetime curvature fluctuations}

An application\cite{BliMa} of the above baby quantum gravity theory is the computation of  the vacuum energy from Planck-scale spacetime curvature fluctuations. This was discussed in the 1950s by J.A. Wheeler as something quantum gravity {\em should} be able tell us, and which he meanwhile sought to guess by analogy with electromagnetism. We actually find a very different answer. 

The first step is to write the Ricci scalar $R$ at the four vertices explicitly. This still depends on $q=e^{\imath\theta}$ for a phase parameter $\theta$ in the QLC and we average over this. The result in field momentum variables is\cite{BliMa}
\[ R(00)=\left({2 k\over 1-k}\right)\left({1\over l_0(1+l)}\right),\quad R(11)=-\left({2 k\over 1+k}\right)\left({1\over l_0(1-l)}\right), \]
\[ R(01)=-\left({1 \over k_0(1-k)}\right)\left({2l \over 1-l}\right),\quad R(10)=\left({1 \over k_0(1+k)}\right)\left({2l \over 1+l}\right).\]
We are interested in $R_{\rm av}$, the average of these four values, as what might be observed at larger scales. The calculations are in \cite{BliMa} and the result there is
\[ \<R_{\rm av}\>\sim {9\imath \over 8 L}\left({1\over  \bar L} - { (144\sqrt{\pi}+35)\over 280\sqrt{\pi}}{1\over \bar L^{3\over 2}}+\cdots\right)\]
in an asymptotic expansion for large $\bar L$. This is imaginary, but that should not come as a surprise given that the scalar curvature defines the propagator for the theory (i.e. from the form of the action). Hence, for a real measure of the curvature from quantum gravity we proposed in \cite{BliMa} to look at its square, namely 
\[ \<R_{\rm av}^2\>\sim {9\over 2^6\sqrt{\pi}\eps L }\left({1\over \bar L^{3\over 2}}+ {9\over 2^{4}  \bar L^{5\over 2}}+\cdots \right)\] 
where  we see the need for the $UV$ regulator $\eps$ that represents the minimum average size of the squares that we are quantising over. If $\eps^2 << L  G$ then the leading term of $\<R_{\rm av}^2\>$ dominates over that of $\<R_{\rm av}\>^2$ and we can think of this equally as measuring the curvature uncertainty\cite{BliMa}
\begin{equation}\label{DelR} \Delta R_{av} \sim {3 G^{3\over 4}\over 8 \pi^{1\over 4} \eps^{1\over 2} L^{5\over 4}}.\end{equation}
This reviews the calculation in \cite{BliMa} and is the main take-away for the next section. Note that $L,\eps, G$ have units of area where until now we have set $c=\hbar=1$. Then $\Delta R_{\rm av}$ has units of inverse area as expected for curvature.

\begin{problem}\rm Have a look at $\Delta R(00),\cdots,\Delta R(11)$ and their average, or better, their root mean square,  in case it behave differently from $\Delta R_{av}$. Also redo both using the Liouville measure for the field variables as in Section~\ref{sec2}.  \end{problem}

\subsection{Carlip-Unruh-Wang explanation of the cosmological constant} 

So far, we have worked within a baby quantum gravity model. Now suppose for purposes of discussion that every region of spacetime at the Planck scale has quantum fluctuations typified by this model. We do not claim exactly this, for one thing our model has a 2D plaquette not a hypercube, but {\em an} answer is better than no answer. In this case, following the spirit if not the actuality of J.A. Wheeler's ideas, it was proposed in \cite{BliMa} that these fluctuations correspond to an gravitational energy density
\[ \rho_{QG}= {\Delta R_{\rm av}\over 8 \pi G }= {3 \over 64 \pi^{5\over 4} G^{1\over 4}\eps^{1\over 2}   L^{5\over 4}}, \]
where we stick with $c=\hbar=1$. Note that if we use a coordinate basis with $\extd x^\mu$ in units of length then $g_{\mu\nu}$ as it occurs in the Einstein tensor is dimensionless. As in \cite{BliMa}, we take $\eps=\lambda_p^2$ the Planck area as the smallest average size of plaquettes in the quantisation. We let $L=\lambda_I^2$ for the typical length of the plaquette (it is more precisely the IR regulator but renormalisation will match it to an observed value of $\<a_{00}\>=\<a_{01}\>\sim 0.6 L$ for larger $\bar L$ and we ignore the order 1 factor here). Then
\[ \rho_{QG}= {3  \over 64 \pi^{5\over 4}} \left({\lambda_p\over\lambda_I}\right)^{5\over 2}\rho_p,\]
where $\rho_p=1/\lambda_p^4=5\times 10^{93}$ g/cm$^3$ is the Planck density. If as in \cite{BliMa}, we take $\lambda_I=2\times 10^{-19}$ (corresponding to particles of energy the order of 100's of TeV) then we arrive at
\[ \rho_{QG}= 6.4 \times 10^{-38}\rho_p= 3\times 10^{56} {\rm g/cm}^3\]
This is still an enormous density and not the observed value $5\times 10^{-30}$ g/cm${}^3$ of dark energy  needed to match the expansion of the Universe in the standard cosmological model. The problem of the cosmological constant is to give a theoretical explanation of this very low but yet not zero density. If one wanted to reverse engineer our calculation and land on $\rho_{QG}$ equal to this, we would need $\lambda_I=2\times 10^{16}$ cm or about 1000 A.U., which is not at all what have in mind in this section but which could nevertheless be interesting in some other context. The $\lambda_I$ we used is chosen to be much smaller than the standard model scale for particle physics so that what we see even at elementary particle scales is an overall smoothed energy density rather than direct observation of quantum gravity fluctuations at the plaquette scale.

This large $\rho_{QG}$ is suitable to feed into the Carlip-Unruh-Wang explanation of the small value of the cosmological constant\cite{Car,WanUn} as follows. The idea is to include such a gravitational energy density into a local form of the Friedmann equation
\[ {\ddot a\over a}=-{4\pi G\over 3}(-2 \rho_{SM}+ (1+3 w)\rho_{QG}),\]
where $\rho_{SM}$ is the density attributable to matter and $w$ models the unknown nature of the gravitational fluctuations,  except that $w>-1/3$ because they are always attractive. Here $a$ is a local expansion parameter and we see that it will be highly oscillatory due to the extremely high value of $\rho_{QG}$ exceeding anything that might come from the $\rho_{SM}$ side. Paradoxially, these oscillations are so rapid that we could never see them as they would cancel out at ordinary scales, so we would see a zero effective cosmological constant up to much smaller `parametric resonance' effects. The argument does not explain the approx $5\times 10^{-30}$ g/cm$^3$ observed density but argues for zero plus corrections. More discussion about how our calculation could support the idea is in \cite{BliMa}. Note that in the absence of a calculation, \cite{Car,WanUn} used an heuristic value $\Delta R= \lambda_p/\lambda_I^3$ motivated by Wheelers `spacetime foam'. This gives
\[ \rho^{\rm Wheeler}_{QG}= {\rho_p\over 8\pi}\left({\lambda_p\over\lambda_I}\right)^3= 10^{50} {\rm g/cm}^3, \]
which still does the trick, but is based on guesswork. Our value is even better for the purpose in hand but more importantly, it comes from an actual model.

\subsection{Speculations for landing directly on the cosmological constant}\label{sec3.4}

A slightly different but not unrelated idea is that the vacuum energy corresponding to the cosmological constant might arise directly from quantum spacetime as a quantum gravity correction to geometry (and hence of Einstein's equation compared to the classical version), i.e. it could be conceptually zero in the QRG but when expanded out in terms of ordinary geometry plus corrections, it might not appear as zero\cite{MaTao}. This requires us to have a better understanding of how the noncommutative variables relate to observations (see Section~\ref{sec6}). Meanwhile, the result in \cite{MaTao} is an example of a deformed Minkowski spacetime algebra where the calculus does not admit a flat metric, hence only curved metrics (the Bertotti-Robinson metric in the case of \cite{MaTao}) could arise from a quantum metric in the classical limit. Key is the commutativity (\ref{gcen}) which is highly constraining for the chosen calculus in the noncommutative case.  In the precent case it forces one to a metric which, in the classical limit, can be seen as coming from a cosmological constant and/or an electromagnetic field. 

Another concrete example where the cosmological constant could be viewed as arising directly from the noncommutative geometry is in context of  Euclideanised 2+1 quantum gravity, where the relevant model quantum spacetime without cosmological constant is $U(su_2)$ regarded as a fuzzy $\R^3$ as in \cite{Hoo}. The role of Poincar\'e quantum group is the quantum double $D(U(su_2))$ and the whole picture is known to q-deform to a $U_q(su_2)$ model spacetime with $D(U_q(su_2))$ for the quantum group symmetry. The latter leads to a Turaev-Viro invariant which underlies the TQFT. The value of $q$ based on dimensional analysis is generally regarded as $q=e^{\lambda_p/\lambda_c}$ where $\lambda_c=1/\sqrt{|\Lambda|}$ and $\Lambda$ is cosmological constant, see \cite{MaSch} for an overview. Here the initial model spacetime is already noncommutative but the momentum space is classical but curved (namely $SU_2$) and the introduction of the cosmological constant makes the momentum space coordinate algebra also $q$-deformed and hence noncommutative, namely to the quantum group $\C_q(SU_2)$. It is also worth noting that so long as $q\ne 1$, the quantum spacetime $U_q(su_2)$ as an algebra is more or less (up to a localisation) the coordinate algebra $B_q(SU_2)$ of the braided version\cite{Ma} of $\C_q(SU_2)$. With the generators of this it actually looks like a q-deformed unit-hyperboloid in q-Minkowski space\cite{CWS,Ma:euc,Ma}. Its QRG, however, remains to be properly understood. 

Meanwhile, a more physical route could be to return to the idea of the preceding section and see if $\rho_{QG}$ of the size needed for the cosmological constant could arise directly as the energy of metric fluctuations measured by $\Delta R$ for quantum gravity on the a noncommutative spacetime, now being viewed as a global model of spacetime not just locally. Ideally, this should be self-consistent with the quantum gravity effects that make the quantum spacetime noncommutative in the first place (see Section~\ref{sec6}). First we note that the relevant value of $q$ for the underlying CFT for 2+1 quantum gravity with cosmological constant  is actually an $n$-root of unity $q^{2\pi\imath\over n}$, where $n$ is the dual coxeter number + the level of a related affine Kac-Moody Lie algebra. We can combine this with the recent discovery in \cite{ArgMa2} that truncation to a finite spacetime induces $q$-deformation. This work starts with the QRG of the half-lattice-line $\N$ (labelled by the natural numbers starting at 1). Here, the inbound metric at $i\to i+1$ near the boundary vertex 1 has $(i+1)/i$ times the outbound value in order for a QLC to exist. When the half line is truncated to $n$ vertices $\bullet$-$\bullet$-$\cdots$-$\bullet$ then all the formulae remain the same but with integers replaced by $q$-integers (so the metric ratio for example become $(i+1)_q/(i)_q$ where $(m)_q=(q^m-q^{-m})/(q-q^{-1})$ and $q=e^{\imath \pi \over n+1}$). Comparing with the 2+1 quantum gravity case, and not worrying about  a factor 2 related to q-integer conventions, we can roughly say that the cosmological constant corresponds to discretisation according to 
\begin{equation}\label{nlam} n\sim  2\pi {\lambda_c\over \lambda_p}=  4 \times 10^{61}\end{equation}
for the observed $\Lambda=10^{-52}$ metres$^{-2}$. There is not expected to be a physical correspondence with the specific finite lattice interval, but we consider it as indicative of a class of models that might apply. (However, the representations of $U(su_2)$ are labelled by a natural number dimension and this becomes truncated to $n$ values for the more relevant reduced quantum group $u_q(su_2)$ at an $n$-th root of unity, so there could be a representation-theoretic duality relating the models).  We see that, broadly speaking, the cosmological constant could also be visible as a consequence of the QRG in related discrete models. From the way it is computed, we imagine $n$ is the lattice size across one dimension, not the total number of vertices. From this point of view, we would like to be able to compute $\rho_{QG}$ in a model with this kind of discretisation and the value we need is
 \[ \rho_{QG}\sim {\rho_p\over n^2}= {5\times 10^{93}\over (4\times 10^{61})^2}=3 \times 10^{-30}{\rm g/cm}^3, \]
to arrive as essentially the desired value, where $\rho_p$ is the Planck density.  This is due to the way that $\lambda_c$ is determined, in conjunction with (\ref{nlam}). Times $8 \pi G$, we therefore want
\begin{equation}\label{DRaim}  \Delta R\sim { 8\pi G\over n^2} \rho_p \sim  {1\over \lambda_c^2}\end{equation}
in the  QRG up to some order 1 constants that we suppress. Here $G$ is the Newton constant which in $\hbar=c=1$ units we replace as before by $\lambda_p^2$ and also in these units, $\rho_p=1/\lambda_p^4$. 

At present, however, we do not have any models with $n$-fold discretisation and where $\Delta R$ is computed as a function of $n$. Quantum gravity on the finite lattice chain $\bullet$-$\bullet$-$\cdots$-$\bullet$  is not studied beyond the 3-node case in \cite{ArgMa2}, and for $\Z_n$ the QRG is known and $\<R\>$ was studied to some extent but not $\Delta R$ itself. For the  4-pointed star in Section~\ref{sec2}, however, we found (\ref{RDtotinv}) that $\Delta R$ is essentially the uncertainty of $1/g$ and from this we could guess (as for derivatives) that this may generically be estimated as  
\[ \Delta R\sim \Delta ({1\over g})\sim {(\Delta g)\over \<g\>^2}\sim {(\bar\Delta g)\over L}\]
up to some order 1 constants, where $\<g\>$ is of order $L$ and  for the Liouville measure we actually had $\eps$ rather than $L$ at the end.  In our context, we would take $\eps=\lambda_p^2$ and $L=\lambda_c^2$ as the relevant scales. This idea that the curvature uncertainty can be estimated from the metric uncertainty for some class of models is the only thing we want to take away -- the specific values of $\Delta R(i)$ for the 4-pointed star  are not our guide here but if we did use the plot in Figure~\ref{star}(a) then we would get the desired value (\ref{DRaim}) for $G_0\approx 6$, where we now use $G_0$ for the dimensionless coupling constant for the model.  We also see in the model that $\Delta g/L$ and $L\Delta R$ are indeed comparable for $G_0$ of this order. 

Therefore, for some class of models, we can guess that $\Delta R$ can be estimated from $\Delta g$. On the other hand, for $\Z_n$, it was found in \cite{Ma:are} that $\bar\Delta g\sim 1/n$ and if this is somewhat indicative of other models with $n$ nodes across the diameter, then we can estimate 
\[ \Delta R\sim {1\over n L} ={\lambda_p\over \lambda_c^3}\quad {\rm or}\quad  \Delta R\sim {1\over n \eps}={1\over \lambda_p\lambda_c}\]
up to order 1 constants, according to whether we are guided by the direct measure or the Liouville measure experience from the 4-pointed star. Neither of these is the dependence we want, but if we take their geometric mean then we do indeed land on (\ref{DRaim}).

Thus, while we have not identified a single model with  $n$ steps across the diameter that gives the right dependence on $n$ and hence on $\lambda_c$ on using (\ref{nlam}),  we see that this plausibly could happen for the right model. At the moment we have merged  experience from two different models to get a speculative feel for how the idea could work. One can also look at the fuzzy sphere where $\<R\>$ and $\bar\Delta\cg$ are worked out,  see\cite{Ma:are} for details. In this model, unlike the $\Z_n$ and 4-pointed star models where the coupling constant in the action is dimensionless, here it has dimension length$^4$. The computed value of $\bar\Delta\cg$ is proportional to this so, while one can reverse engineer to get the correct answer,  this requires more justification as to the value of the coupling constant.  It is also the case that the models discussed are Euclidean rather then Lorentzian, which also needs to be addressed. 
 
 A further and more conceptual issue is that if spacetime is noncommutative  globally and the cosmological constant comes from this one way or another, we still need to know how classical gravity, which does not have the rigidities  of the quantum case, is meant to emerge. One route could be to relax the axioms of a quantum metric, but with correspondingly harder calculations. This was avoided in the preceding section, where we only used QRG at the plaquette scale and assumed that lower energy physics (including the standard model) applies at larger scales than this. This is also similar to the next section where we provisionally use quantum geometry not for spacetime (as one eventually should)  but just for an internal fibre geometry at each point of classical spacetime. One could then have the spacetime cosmological constant arising from fluctuations in the internal geometry without (in this approximation) affecting gravity on spacetime.
 
 \begin{problem}\rm Extend \cite{Ma:are} to include  $\Delta R$ for the $\Z_n$ and fuzzy sphere models, with  a view to direct calculation of $\rho_{QM}$ rather than by indirect means. Also consider Lorentzian versions with $\imath$ in the action as well as the finite lattice chain $\bullet$-$\bullet$-$\cdots$-$\bullet$ with the QRG given in \cite{ArgMa2}.
 \end{problem}

\section{Quantum spacetime origin of gravity+Yang Mills}\label{sec4}

Another recent success of the QRG approach is in explaining why at low energies we have gravity + Yang-Mills in the first place. I am lumping electromagnetism along with Yang-Mills here, I just mean any gauge theory. The idea, inspired by pioneering work of Connes and collaborators\cite{Con95} is to revisit an old Kaluza-Klein (KK) idea that what we observe is actually gravity on an extend spacetime with a compact internal fibre at each point. This idea in its original form never really worked  for three reasons\cite{ArgMa3,LiuMa2,LiuMa3,LiuMa4,LiuMa5}:

\begin{enumerate} 
\item[(i)] The metric needs to be of a very special `cylinder ansatz' form to recover gravity + Yang Mills\cite{Coq}. So all this is really saying it that gravity on the product has so much freedom that one can find in there the modes we want, not that these are predicted or explained.

\item[(ii)] To work, the geometry on the fibre has to be again artificially set to be of constant size, otherwise the Yang-Mills coupling constant will vary in spacetime. The size is governed by a Liouville-like field but the classical equations of motion for this are {\em not} compatible with this field being constant.

\item[(iii)] This constant radius is $23\lambda_p$ for a circle in the case of electromagnetism  or $11\lambda_p$ for the $S^3$ in the case of electroweak Yang-Mills, for example.\end{enumerate}

The last of these was not considered a problem at the time, but if the size is of Planck scale order then this suggests that we should not have a classical spacetime fibre in the first place but a noncommutative one!  This then turns out to solve the first two complaints \cite{ArgMa3,LiuMa2,LiuMa3, LiuMa5} as follows. Which it turn says that the structure of gravity + Yang-Mills that we see comes out of gravity on a product spacetime in conjunction with quantum gravity corrections at least in the fibre dimensions. 

To analyse this situation in QRG, we write $A=C^\infty(M)\tens A_f$ for some fibre noncommutative geometry. We suppose that $\Omega(A)=\Omega(M)\underline{\tens} \Omega(A_f)$ is also a (now, graded) tensor product form for the exterior algebra. We suppose local coordinates on $M$ and for $\Omega(A_f)$ we assume a basis $e^i$ of $\Omega^1$ over $A_f$ and that (i) the $e^i$ are central (ii) the algebra $A_f$ has trivial centre (i.e. is sufficiently noncommutative). Then a quantum metric on the product {\em must} have the form
\[ \cg=g_{\mu\nu}\extd x^\mu\tens \extd x^\nu + A_{\mu,i}(\extd x^\mu\tens e^i + e^i\tens \extd x^\mu) + h_{ij}e^i\tens e^j \]
{\em where $g_{\mu\nu},A_\mu,h_{ij}\in C^\infty(M)$ do not depend on the fibre}. This comes directly from  (\ref{gcen}) combined with the assumptions stated. We also impose symmetry in the cross term, symmetry of $g_{\mu\nu}$ and an appropriate quantum symmetry for the fibre metric $\cg_f=h_{ij}e^i\tens e^j$. That the coefficients here depend only on spacetime is the cylinder ansatz (i) needed (in the case of classical fibre) in KK theory, now not as an ansatz but following from the rigidity of QRG expressed in (\ref{gcen}). 

We then make a further assumption regarding the QRG, namely that the generalised braiding of the QLC on the product spacetime is the flip map when any argument is spacetime $\extd x^\mu$. Then there exists a unique QLC on the product. This was shown in the  concrete case of $A_f$ a matrix algebra $M_2(\C)$ in \cite{LiuMa2} or $A_f$ a fuzzy sphere in \cite{LiuMa3,LiuMa4} each with standard differential structures, but the same  result is expected generally so long as $\cg_f$ admits a unique (or natural choice of) fibre QLC\cite{ArgMa3}. We omit the details of the QLC,  but once found, we crank the QRG handle and compute the Ricci tensor with respect to a natural lifting map $i$ on $\Omega^2(A_f)$ and the obvious antisymmetric lift otherwise. For the fuzzy sphere, the $e^i$ are Grassmann algebras under the wedge product and we just lift $i(e^i\wedge e^j)=(e^i\tens e^j-e^j\tens e^i)/2$ as usual for quantum gravity on the fuzzy sphere\cite{Ma:are}.  After a very long calculation, the result, stated here in the fuzzy sphere case, is that the Ricci scalar on the product splits as \cite{LiuMa3}
\begin{align*}
\ R=&\tilde R_M+R_h+\frac{1}{8}h_{ij}\tilde F^i_{\mu\nu}\tilde F^{j\mu\nu}+\frac{1}{2}\tilde\nabla^\alpha {\rm Tr}(\Phi_\alpha)+\frac{1}{8}\big({\rm Tr}(\Phi_\alpha\Phi^\alpha)+{\rm Tr}(\Phi_\alpha){\rm Tr}(\Phi^\alpha)\big),
\end{align*}
where $\tilde R_M$ in QRG conventions is $-1/2$ of the usual Ricci curvature of the metric $\tilde g$ and $\tilde F$ the usual curvature of the gauge field $\tilde A^i_\mu$, where
\[ \tilde g_{\mu\nu}=g_{\mu\nu}-h^{ij}A_{\mu i}A_{\nu j},\quad \tilde A^i_\mu= h^{ij}A_{\mu,i}. \]
We use $\tilde g$ for the raising the indices for the YM actions. We also have 
\[R_{h}={e^{-{\rm Tr}(\Phi)}\over 2}\Big({\rm Tr}(e^{2\Phi})-{1\over 2}{\rm Tr}(e^\Phi)^2\Big) \]
 the Ricci scalar on the fuzzy sphere as previously used for quantum gravity there, now regarded as a potential term for a Liouville-type field
\[ \Phi:=\ln(\underline h),\]
where we take the log of the positive matrix $\underline h=\{h_{ij}\}$. Finally, we have used the notation
\[\Phi^i_{\alpha j}:=h^{ik}\tilde\nabla_{\alpha A} h_{kj}\]
using the covariant derivative with respect to $\tilde g$ and $\tilde A_{\mu i}$ acting on lower latin index tensors by $\eps_{ijk}$. Actually, these formulae are essentially the same as one gets from usual KK theory with classical $S^3$ fibre\cite{Coq}  but derived very differently. We see that to get Yang-Mills, we need $h_{ij}$ to be constant. 

This leaves us with the problem (ii). Our approach to this proposed in \cite{LiuMa5} is that in fact, the fibre being quantum, one cannot use normal variational calculus to derive the equations of motion for the Liouville-type field. In the absence of a theory for this (see the discussion in Section~\ref{sec6}) we prosed an alternative which is to quantise the Liouville field so as to obtain an effective theory for the remaining fields $\tilde g, \tilde A_\mu$. This is done at each point in spacetime as quantum gravity on the fibre fuzzy sphere, in the present case. At first sight it seems clear that to get the right answer, we need $\<h_{ij}\>=h\delta_{ij}$ for $h$ a constant and of suitable value so as to obtain the desired Yang-Mills coupling constant. To cut a long story short, we can expect (and get) that the result is proportional to $\delta_{ij}$ provided we can do the quantum gravity in a way that preserves the rotational symmetry of the fuzzy sphere. And we can get the desired value {\em because} the quantum gravity theory is divergent and has to be renormalised. The renormalisation involves assigning $h$ in the expectation value at some scale and running from there. But as each theory is done at each point of spacetime independently, we have enough freedom to {\em choose} the same value everywhere. This also boils down to the effective values of the derivatives $\Phi_\alpha$ being ignored. There are some critical assumptions here, so this is not exactly a derivation but rather a plausible scenario. 

This scenario is explored further in \cite{LiuMa5} and here we mention only a few highlights. First, the action on the product spacetime given $R$ above requires an integration
\[ \int_A=\int_M\extd^4x \sqrt{-\tilde g} \int_{A_f},\quad   V_f=\int_{A_f} 1\]
for some positive linear functional $\int_{A_f}: A_f\to \C$ in the role of integration, except that we only need to specify $V_f$ since all functions in $R$ do not depend on the fuzzy sphere coordinates. The natural choice used in previous work for quantum gravity on the fuzzy sphere is $V_f=\det(\underline h)$. 

We also need an $\imath$ in front of the action as the spacetime part from $\tilde R_M$ has to be a Lorentzian gravity theory.  But this implies that the action from $R_h$ for quantum gravity on the fibre also has an inherited $\imath$, which means a Lorentzian\cite{LiuMa5} as opposed to previous Euclidean quantum gravity as in \cite{Ma:are}. This is the first complication. The second is that suppose we have a coupling constant $G_0$ for the overall product action. The effective gravitational and fibre quantum gravity coupling constants are then 
\[ G_N={ G_0\over \<V_f\>},\quad G_f={G_0\over V_M},\]
where $V_M$ is the volume of spacetime, but we can use any sufficiently large box or relevant scale for example $\lambda_c^4$. Likewise the thing that enters the Yang-Mills part of the action is not exactly $\<h_{ij}\>$, but $\<V_f h_{ij}\>$. Hence it could be argued that it is really this which needs to be proportional to $\delta_{ij}$, and the constant of proportionality divided by $G_0$ has to match what we need for the Yang-Mills action with its coupling constant. Both scenarios, (a) using $\<h_{ij}\>$ and its determinant for $V_f$, and  (b) using $\<V_f\>$ and $\<V_f h_{ij}\>$, are explored in \cite{LiuMa5} and also three different choices for the measure of integration on the moduli of quantum metrics on the fibre. As in previous work, it is enough to work with diagonal $h_{ij}={\rm diag}(\lambda_1,\lambda_2,\lambda_3)$ or $\phi_i=\ln(\lambda_i)$ for three Liouville-like fields. One `geometric' measure is to regard $\underline h$ as an element of the symmetric space of positive matrices, another is the `Liouville' measure $\extd^3\phi$, and we can also directly use $\extd^3\lambda$. We then proceeded to match to the actual values of $G_N$ and Yang-Mills coupling for the different scenarios and the main take-away's are\cite{LiuMa5}: 

\begin{itemize}
\item Some combinations of scenario and measure are viable and some are not, in particular the more logical (b) scenario with the Liouville measure works well. The geometric measure has similar behaviour but was studied in less detail. 
\item We need to be in the extremely low coupling constant regime of quantum gravity on the fibre. 
\item The IR cutoff length scale $\sqrt{L}$ needs to be around 150-200 $\lambda_p$ which is a viable dynamic range for a theory of quantum gravity. 
\item The relevant expectation values are not necessarily real and we just used their absolute values in calculations, but this could suggest  a small imaginary component in Yang-Mills theory.
\end{itemize}

It is fair to say that this first work \cite{LiuMa5} needs plenty of refinement. Hence, the best we can say at the moment is that this approach is promising, but needs more work. The second item here, in particular, needs a new analytic approach because the numerical integrations were unstable for small coupling and we had to extrapolate downward from the values that could be reached. Finally, of course, these details were for the fuzzy sphere and everything could be tried for other QRG's $A_f$.

\begin{problem}\rm (a) Repeat the QRG version of KK theory for $A_f$ some specific algebras suggested in Connes approach to the Standard Model\cite{Con95} (such as two copies of quaternions), suitably re-interpreted. Using for $A_f$ the octonions as a quasi-associative geometry could also be of interest. Also of interest could be  $A_f=U(\cg_7)$ where $\cg_7$ is the 7-dimensional Malcev algebra tangent to the 7-sphere. This is not a Lie algebra and $A_f$ is again not associative, but is a Hopf quasi-algebra\cite{KliMa}. This replaces fuzzy $\R^3$ and could also be of interest as a non-associative quantum spacetime in its own right. (b) Analyse quantum gravity on the fibre analytically in the very low coupling regime. \end{problem}

\section{Wave functions on spacetime and generally covariant quantum mechanics}\label{sec5}

Another recent physical application of QRG is a theory with Edwin Beggs of {\em generally covariant quantum mechanics}\cite{BegMa:qm}. It arose out of a concept of quantum geodesics applied to the noncommutative algebra $\CD(M)$ of differential operators on a manifold. The latter is the coordinate-invariant notion of the Heisenberg algebra in any local coordinate chart and the quantum geodesic flow gives evolution equations which quantise geodesic motion in a Heisenberg picture of the evolution of operators. I will return to this later, but meanwhile it turns out that the corresponding Schr\"odinger picture for the evolution of states is completely accessible without knowing any noncommutative geometry. Namely, it comes down to `Klein-Gordon quantum mechanics' or the {\em Klein-Gordon flow}\cite{BegMa:qm,BegMa:flrw,KumMa}
\begin{equation}\label{KGflow}
-\mathrm{i} \frac{\partial \varphi}{\partial s} = \frac{\hbar}{2m} \square \varphi,
\end{equation}
where \(\square\) is the Klein-Gordon operator. Here, $m$ is a parameter of mass dimensions but it is {\em not} the mass of any one Klein-Gordon field. Rather, fields $\varphi$ evolve with respect to a parameter $s$ and on-shell fields are merely stationary modes for the evolution, more analogous to atomic states in time-independent quantum mechanics. At each $s$, $\varphi$ is viewed as a wave function in analogy with quantum mechanics {\em but over spacetime}. While easy enough to write down, what wave-functions on spacetime mean operationally is unclear (who is doing the quantum measurement?) and related to this, who's time corresponds to the parameter $s$ with respect to which the fields are evolving? Our approach is to not get too hung-up on the interpretation, which could emerge with time and experience with examples, but for now to be guided by the mathematics. We just note that  the right hand side is generally covariant, so on the left hand side, $s$ should be some kind of collective proper time and this further justified later at the Heisenberg picture level. Also note that since we do not work with on-shell fields, the theory has some of the information normally approached through quantum field theory, without being quantum field theory. This turns out to be a very calculable paradigm with hints at lots of interesting physics even if the full interpretation is not yet known. 

\subsection{Pseudo-quantum mechanics on a static spacetime, horizon modes and entropy}

In particular, I will recap from \cite{BegMa:qm} for the case of a Schwarzschild black hole. Similar results were recently computed in Kruskal-Szekeres coordinates in \cite{KumMa} confirming that the findings are indeed not coordinate artefacts, and also allowing one to now get inside the black-hole. Working in standard Schwarzschild coordinates where the metric is
\[  \cg=-(1-{r_s\over r})\extd t^2 + {1\over 1-{r_s\over r}}\extd r^2 + r^2(\extd\theta^2 + \sin^2(\theta) \extd\phi^2),\]
we look at the evolution with respect to $\varphi$ of the form
\begin{equation}\label{varphi} \varphi(t,r,\theta,\phi)= e^{p_t\over\lambda} \psi(r,\theta,\phi),\end{equation}
where $\psi$ at each $s$ is now a wave function in three dimensions (independent of $t$). In this case the Klein-Gordon flow reduces to
\begin{equation}\label{psqmbh} -\lambda{\del\psi\over\del s}=( -{\hbar^2\over 2m}\Delta + V_{eff})\psi,\quad V_{eff}=-(1-{r_s\over r})^{-1}{p_t^2\over 2m},\end{equation}
\[ \Delta:=(1-{r_s\over r}){\del^2\over\del r^2}+{1\over r}(2-{r_s\over r}){\del\over\del_r}+{1\over r^2}\left( {\del^2\over\del\theta^2} + {1\over\sin^2(\theta)}{\del^2\over\del\phi^2}+ \cot(\theta){\del\over\del\theta}\right), \]
which now looks a lot like regular quantum mechanics except that $s$ plays the role of time not $t$. The special case of a Klein-Gordon flow where we factor the $t$-dependence like this is called {\em pseudo-quantum mechanics}, applied here in the case of a static spacetime but also applicable with modifications in other cases such as \cite{BegMa:flrw}, where there is a natural time coordinate that we swap out in favour of $s$. At this point, we can now use the tools of regular quantum mechanics including completing to an $L^2$ space of such fields $\psi$, except that this is now with respect to the inherited $\sqrt{-g}$ measure from spacetime, in our case 
\[ ||\psi||_{L^2}= \int_{r_s}^\infty r^2 \extd r\int_0^\pi |\sin(\theta)|\extd\theta \int_0^{2\pi}\extd\phi\,   \bar\psi \psi, \]
where we focus as in \cite{BegMa:qm}  on the exterior region. Moreover, one can show that the norm of $\psi$ does not change under the evolution under $s$, hence we can take a probabilistic interpretation where $|\psi|^2$ is a probability density over space with respect to the above measure of integration. As in quantum mechanics, we can solve (\ref{psqmbh}) both for time-dependent solutions from an initial $\psi$ as $s=0$, and for time-independent stationary modes where
\[ ( -{\hbar^2\over 2m}\Delta + V_{eff}+ {p_t^2\over 2m})\psi=E\psi,\quad   -p_t= \sqrt{m_{KG}^2+ 2m E}.\]
We have subtracted off the rest energy to match conventions in regular quantum mechanics, with $m_{KG}$  the Klein-Gordon mass of the corresponding $\varphi$ in the sense $\square\varphi= {m^2_{KG}\over\hbar^2}\varphi$. Also, $-p_t=\hbar\omega$ in terms of the corresponding frequency. 

We refer to \cite{BegMa:qm} for details of both types of solution. This focussed mainly on $\psi(r)$, where there is no $\theta,\phi$ dependence, and the key findings for the evolution solutions are:

\begin{itemize}\item A real Gaussian bump $\psi$ evolves into a complex wave with $|\psi|^2$ a bump that dissipates over time. The same happens in flat spacetime\cite{BegMa:geo}, so this is expected.
\item When the region of disturbance approaches the horizon, it generates `horizon modes' there which (in principle) have infinitely small wavelength approaching it.
\item After a period of time, the original bump is entirely turned into horizon modes.
\item The classical entropy of the probability density, i.e. 
\[ S(\psi)=\< -\ln(|\psi|^2\>=-\int_{r_s}^\infty \extd r \, r^2 |\psi|^2 \ln(|\psi|^2)\]
for a normalised state increases throughout this process. This appears (based on random trials) to be the case for all initial real positive $\psi(0)=\sqrt{\rho(0)}$ for any localised density $\rho(0)$, or any constant phase times this.
\item The expected position 
\[ \< r\>=\int_{r_s}^\infty \extd r \,  r^3 |\psi|^2 \]
for a normalised state also increases throughout the process.
\end{itemize}
The entropy increasing is one of the things that suggests that this could be a physical process of some sort, in which an initial Gaussian bump is swallowed by the black-hole, releasing horizon modes. That $\<r\>$ increases is consistent with the expected Ehrenfest theorem and arises because the horizon modes propagate away from the horizon. These are results in \cite{BegMa:qm}.

\begin{figure}
\[ \includegraphics[scale=0.75]{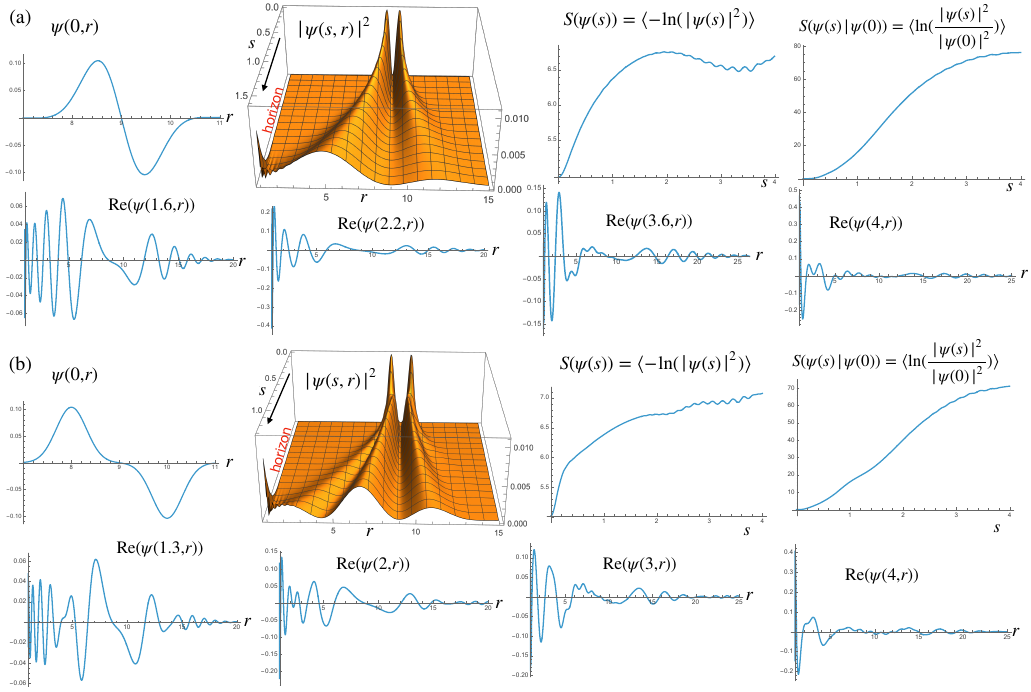} \]
\caption{(a) Horizon modes generated at the horizon when an initial real dipole is swallowed by the black-hole can lower the classical entropy $S(\psi(s)))$ but (b) not when the two parts of the initial dipole are sufficiently separated so as to not interfere. The classical entropy  relative to the initial $\psi(0)$ nevertheless increases. \label{hor}}
\end{figure}

As a new result in the present work, we now look more closely at the classical entropy in an example where it {\em decreases}, see Figure~\ref{hor}(a), namely  for a a real dipole $\psi$. This is again swallowed by the black-hole, starting at about $s=1.3$ with the creation of horizon modes which eventually are all that remain pushed up at the horizon.  We work in units where $r_s=1$ and for practical reasons we cut off the domain at $1+\delta$ where $\delta=0.0001$ and we also set the resolution scale of the numerical solver to be this (i.e. we take this for MaxCellMeasure for NDSolve in Mathematica). The resolution limits the scale of oscillations that we can see. The work \cite{KumMa} studies dependence on both the domain cutoff and resolution in detail for the 1-bump and the full story appears to be similar for the dipole here.  We also now add the relative entropy
\[ S(\psi(s)|\psi(0))= \<\ln ({\psi(s)\over\psi(0)})\>\]
of the classical density at time $s$ compared to the intiial one (this is the Kullback-Leibler divergence\cite{KL} between the states in information geometry). In both cases, the expectation value is computed with probability density $|\psi(s)|^2$ with respect to the measure and we have added $10^{-40}$ to all instances of $|\psi|^2$ inside the logarithm to prevent numerical overflows where these are too close to zero (nothing changes if we  instead use, say, $10^{-50}$). We see that after about $s=2.2$, the entropy starts to decrease for a bit until the entire dipole has been fully swallowed by $s=3.6$, while the relative entropy increases throughout. Part (b) shows that this can be attributed to the non-classical effect of interference between the positive and negative parts of the initial dipole wave-function, because it disappears when these are separated so as to not significantly overlap. Indeed, if an initial state is a sum $\psi(0)=\sum_i \psi_i(0)$ where the $\psi_i(0)$ do not overlap and where each has a constant phase, then $S(\psi(s))$ is a convex linear combination of the $S(\psi_i(s))$. Where the latter are each increasing by the apparent result in\cite{BegMa:qm}, it follows that so is $S(\psi(s))$. What this suggests is that the classical entropy is a good measure when the system is effectively classical with no quantum mechanical self-interference between the different regions of support, but that there is a quantum interference component to the entropy still missing and to be accounted for. This also implies that the ripples in $S(\psi(s))$ at later $s$, at least in the case (b), are  numerical artefacts although they remain to be fully understood,  being well above the numerical resolution scale. There are also subtle differences in the structure of $\psi(s)$ between the two cases, which may be a clue. For example, the normal process for one bump is that the horizon modes increase in height and the density gets increasing located near the horizon, but this does not happen during the period where the classical entropy decreases.  For both parts of the figure, integrity of the numerics was verified by checking that $|\psi(s)|^2$ is constant during the evolution up to numerical errors of $<0.9\%$ over the range plotted,  and that these decrease if the everything is done at higher resolution (smaller $\delta$).

For the time-independent solutions around a black-hole, the potential $V_{eff}$ is close to that of an atom and one has similar normalisable solutions in the exterior when $m^2_{KG}>{p_t^2}$, see \cite{BegMa:qm}. These are again increasingly oscillatory in a fractal-like manner approaching the horizon. But because of this, we cannot usefully set boundary conditions at the horizon, nor could we get inside the black hole in these coordinates. As a result the `atomic spectrum' of exterior gravatom modes (where the black-hole plays the role of nucleus) in \cite{BegMa:qm} was continuous and not discrete. In the sequel \cite{KumMa}, however, we use Kruskal-Szekeres coordinates  and can look just as easily inside the black-hole. Key new findings are \cite{KumMa}:

\begin{itemize}
\item A Gaussian bump inside again dissipates and generates horizon modes now on the inside of the horizon and propagating away from it (towards the singularity). The entropy for a single bump increases as long as horizon modes do not reach the singularity and the expected value of $z=UV$ also moves towards the singularity located at $z=1$. 
\item There are atom-like modes inside the black-hole with fractal behaviour approaching the horizon from the inside. Moreover, there is a natural mixed bounday condition at the singularity, in which case the atomic spectrum of these modes for fixed energy $p_t$, i.e. the Klein-Gordon mass spectrum of such solutions of the Klein-Gordon equations inside the black-hole, is discrete.
\end{itemize}

\begin{figure}
\[ \includegraphics[scale=0.9]{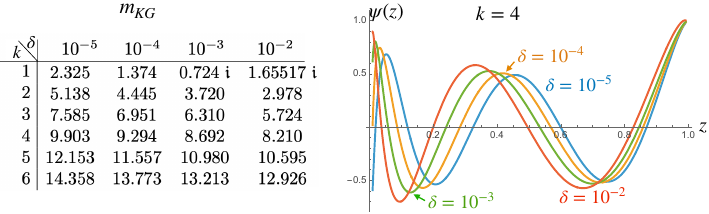} \]
\caption{Mass spectrum $m_{KG}$ from \cite{KumMa} of `atomic' solutions of the Klein-Gordon equations inside the black-hole in units of $\hbar/r_s$ for different value of domain cutoff $\delta=z_{min}/r_s$ and for $p_t=-\hbar/r_s$. Here $k$ is the number of main zero-crossings in $\psi$ away from the horizon as shown for $k=4$. \label{mkg} }
\end{figure}


As before, even in Kruskal-Szekeres coordinates, the equations being solved are singular at the horizon and we have to cut off, now  just inside the horizon located at $z=0$. The values of the discrete spectrum depend on this cut-off of the domain at $z_{min}=\delta$,  but not strongly, as shown by the table in Figure~\ref{mkg}. The figure also shows the eigenfunctions corresponding to the $k=4$ row of the table (these have 4 zero-crossings away from the horizon) and for $p_t=-1$. More details are in \cite{KumMa}, but in first approximation one can see that
\[ m_{KG}\sim 3.6 k\]
in units of $\hbar/r_s$ i.e. in units of the mass of a particle whose Compton wavelength was the Schwarzschild radius, after we put back in the units. The results shown are for $p_t=-\hbar/r_s$ for the $t$-dependence in the actual Klein-Gordon mode $\varphi$ in (\ref{varphi}) once we put back in $\hbar, r_s$ (where the actual calculations in \cite{KumMa} are done in units with $r_s=\hbar=1$). Also of interest is that \cite{KumMa}  actually solves for $E_{KG}=-{m_{KG}^2\over 2m}$ as the eigenvalue of the original Klein-Gordon flow (\ref{KGflow}) for $\varphi$ and, as can be seen in the table, two of the $k=1$ solutions are positive for this, i.e. tachyonic for mass. The $m_{KG}=2.978$ mode at $k=2$ for example, as $\delta$ decreases, morphs to the $k=1$ mode at $m_{KG}=0.724\, \imath$ and on the way passes at $\delta= 1.227\times 10^{-3}$ through a massless mode where $m_{KG}=0$. 

So far, $\delta$ is a parameter, but stepping back, one of the principal take-aways from this and other detailed analysis in \cite{KumMa} is that numerical integration smooths the infinitely-fast oscillation modes at the horizon by cutting them off at the solver resolution scale. The latter is in principle a different parameter from the domain cutoff $\delta$, but the smallest and hence most natural choice is to make the resolution scale it equal to it.  This then allows, for example, to solve the evolution of an interior or exterior Gaussian bump through the horizon (where it is found to extend as more horizon modes in a thin layer on the other side of the horizon). Whilst this is a numerical artefact, it is similar to what one might expect from a discrete geometry or from certain types of noncommutative geometry where modes of wavelength smaller than Planck scale are naturally cut off. In other words, it  is proposed in \cite{KumMa} that:

\begin{itemize}\item While it should be seen only as indicative, the finite resolution of the numerical PDE solver is a kind of poor man's quantum geometry indicative of quantum gravity corrections to the geometry in the role of the Planck scale (or rather in our context  $\lambda_p/r_s$ with $r_s$ the Schwarzschild radius). 
\end{itemize}

As an example, if we set  $\delta=\lambda_p/r_s$ as the resolution scale then there is a particular size of black-hole (about  $800 \lambda_p$, i.e. around the GUT scale) where the $k=2$ mode at $\delta=10^{-2}$ in the table becomes massless. While speculative, the numerical work therefore suggests that black-holes the size of GUT scale elementary particles should have quantum gravity corrections in their Klein-Gordon spectrum of fixed-energy modes inside the black-hole. This is intriguing, even if we are a long way from understanding a physically relevant scenario for such black-holes, but see \cite{grav}. There may also be similar effects outside the black-hole. Finally, we note that the Klein-Gordon flow can also be useful in the FLRW cosmological model. This is not static but one nevertheless has a comparable analysis using Bessel functions and novel behaviour at exactly the range of Hubble constant scales used in inflation models, see \cite{BegMa:flrw}. This work also finds gravitationally bound, but now cosmological, solutions of the Klein-Gordon equations. 

\begin{problem}\rm (a)  Develop a notion of entropy of a wave function that accounts for self-interference and always increases when the initial wave function is swallowed by a black-hole.  (b) Relate the fractal behaviour of the horizon modes at the horizon and their numerical resolution across it to the `quantum skin' at the horizon in a wave operator approach to noncommutative black-holes\cite[Sec 9.3]{BegMa}. (c) Look at such phenomena for other curved geometries. 
\end{problem}

\subsection{Heisenberg picture and nonassociative QRG}\label{grqm}

Related to the preceding section is an approach to quantum geodesic flows  on the noncommutative algebra $\CD(M)$ of differential operators\cite{BegMa:geo, BegMa:qm,BegMa:flrw} viewed as  a coordinate-invariant version of the Heisenberg algebra. I will briefly recap this here, because it helps to justify $s$ as a proper time parameter operator-level due to the resulting `Heisenberg picture' geodesic flow equations
\begin{align}
m \frac{\mathrm{d} x^\mu}{\mathrm{d} s} &= g^{\mu \nu} p_\nu - \frac{\lambda}{2} \Gamma^\mu, \label{momenta} \\
m \frac{\mathrm{d} p_\mu}{\mathrm{d} s} &= \Gamma^\nu_{\mu \sigma} g^{\sigma \rho} \left( p_\nu p_\rho - \lambda \Gamma^\tau_{\nu \rho} p_\tau \right) + \frac{\lambda}{2} g^{\alpha \beta} \Gamma^\nu_{\beta \alpha, \mu} p_\nu, \label{momentaFlow}
\end{align}
written in first order phase space form and in local coordinates where $\lambda=-\imath\hbar$ and $[x^\mu,p_\nu]=\imath\hbar\delta^\mu{}_\nu$. Here  \(g^{\mu \nu}\) is the metric, \(\Gamma^\mu = \Gamma^\mu{}_{\alpha \beta} g^{\alpha \beta}\) are contracted Christoffel symbols, and $m$ is a mass parameter. More precisely, $x^\mu, p^\nu$ here are local generators extendible as global elements of $\CD(M)$, the algebra of differential operators. This is generated  by smooth functions $f$ and vector fields $X$ with cross relations $[X,f]=\lambda X(f)$ and relations $X Y-Y X=\lambda [X,Y]_{Lie}$ in terms of the Lie bracket of two vector fields $X,Y$. If we take the classical limit $\lambda\to 0$ then locally (\ref{momenta})-(\ref{momentaFlow}) become the usual equations for a geodesic on first order form, where $s$ indeed is exactly proper time. At the next level, again locally for convenience, we can represent $x^\mu, p^\nu$ as (unbounded) operators on $L^2(M)$ where $x^\mu$ acts by left multiplication and $p^\mu=-\imath\hbar{\del\over \del_\mu}$ as expected for momentum. Then 
\[ p^2_{tot}=-g_{\mu\nu}(p^\mu p^\nu - \lambda \Gamma^\rho_{\mu\nu} p_\rho)\]
is a conserved quantity and acts as the Laplacian $\square$. 

On the other hand, we do not rule out that these equations could have order $\lambda^2$ corrections when $M$ has curvature. The reason is that they are derived from noncommutative geometry on $\CD(M)$ with a natural calculus constructed from the desired Hamiltonian. Then \cite{BegMa:qm} uses such a calculus to write down a quantum geodesic flow, although not all of the quantum geometry, notably a certain bimodule connection, has been worked out. This connection does  exist in the flat space case\cite{BegMa:geo} and it may be more a matter of computational difficulty than existence in the general case. The most unusual thing about this calculus, however, is that \cite{BegMa:qm}:

\begin{itemize}\item The calculus $\Omega(\CD(M))$ is nonassociative  at order $\lambda^2$ in the presence of background curvature.
\item The relations given by moving $\theta'=\extd s$ in (\ref{momenta})-(\ref{momentaFlow}) to the right hand side kill the  nonassociativity   when we impose them.
\end{itemize}

This represents a new and more algebraic explanation for why things evolve in the first place, different from the principle of least action. The evolution is actually of a Hamiltonian form given by $[\ch, ]$ for $\ch=-p_{tot}^2/(2m)$, and this is why the corresponding Schr\"odinger picture is the Klein-Gordon flow. But what we see is that underlying it is some nontrivial and slightly nonassociative geometry. Meanwhile, the relations (\ref{momenta})-(\ref{momentaFlow}) have been relatively little used so far, e.g. to find an Ehrenfest theorem for motion around a black-hole\cite{BegMa:qm}, and could certainly be looked at more broadly and for other curved spacetimes.

\begin{problem}\rm  Develop the Heisenberg flow picture based on (\ref{momenta})-(\ref{momentaFlow}) further for the black-hole, FLRW model and other curved spacetimes, with a view to seeing if the order $\lambda$ corrections which one might not have naively put into the Hamiltonian have a physical prediction and if there are higher order corrections.
\end{problem}

\subsection{Classical and quantum geodesic flows} 

So far, I have mentioned quantum geodesics but not said what they are. In fact the full conceptual explanation comes from the theory of $A$-$B$-bimodule connections\cite{Beg:geo,BegMa:geo} which is a polarised version of a usual $A$-bimodule connection. Here $A$ is the spacetime coordinate algebra and $B$ is the coordinate algebra of the parameter space, both of which could be noncommutative,  although so far only classical $B=C^\infty(\R)$ (for a single classical time parameter $s$) has been studied in any detail. In this case, the theory is less fancy and amounts to a quantum vector field as a right module map $X: \Omega^1\to A$, a connection on the space of such vector fields (for example induced by a QLC on $\Omega^1$), a positive linear functional or `integration' $\int: A\to \C$ and a  density $\rho=\psi^*\psi$, i.e. a positive element of the $*$-algebra normalised (ideally) so that $\int \rho=1$. Both $X$ and $\rho$ or $\psi$ are subject to certain evolution or flow equations with respect to $s$. Examples with $A$ or its differential calculus noncommutative  have been constructed in \cite{Beg:geo,BegMa:geo,LiuMa1, BegMa:cur,BegMa:gra, BegMa:qm}. 

In this section, I will briefly explain the motivation behind recent work with K. Kumar in \cite{KumMa}, where we look in practice at what quantum geodesics reduce to when applied in the classical case of $A=C^\infty(M)$ for $M$ a smooth (pseudo)-Riemannian manifold such as a black-hole. The general classical limit was the motivation from the start in \cite{Beg:geo}, but we show that it is also of practical interest in its own right as it provides new tools for classical General Relatively. It also provides the necessary intuition for  how to think about quantum geodesics more generally, in particular the full meaning of the collective proper time parameter $s$. 

One thing our classical geodesic flows are {\em not}, is the conventional notion of a geodesic flow in the geometry literature, which is actually a flow on the normal bundle that evolves the point $(x,v)$, where $x\in M$ and $v\in T_xM$ is unit vector, as a geodesic. Classical quantum geodesic flows are not this because they are evolutions of densities $\rho$ directly on $M$ itself. As such, they are closer to relativistic fluid mechanics\cite{Olt}  or optimal transport theory\cite{LotVel}, but still different from both of these. The intuitive picture in our case is to imagine that the tangent vectors of all the dust particles provide a vector field $X$. This turns out\cite{Beg:geo}, however, to obey its own equation that does not even mention $\rho$, namely the {\em geodesic velocity equation},
\begin{equation} \label{velocityflow} 
\dot X + \nabla_X X=0, 
\end{equation}
where the dot denotes $\extd\over\extd s$. Next\cite{Beg:geo}, $X$ relates to the rate of change of  a density $\rho$,  according to the {\em density flow equation},
\begin{equation}\label{densityflow}
\dot\rho + X(\rho) + \rho\, \div(X)=0.
\end{equation}
Note, however, that this equation determines $\dot \rho$ from $X$,  but not the other way around. Thus, the classical limit of the concept of a quantum geodesic tears apart the usual notion of a single geodesic (where the position comes first and its motion determines the velocity) and puts this data back in reverse order, where  $X$ is a field in its own right that then determines the flow of any density $\rho$.  

Therefore the key goal in \cite{KumMa} is to explore the meaning of such geodesic velocity fields $X$ given that they are the more fundamental object. Another goal here is to see what happens if we write $\rho=|\psi|^2$ as in quantum mechanics, where $\psi$ evolves as
\begin{equation}\label{ampflow} 
\dot\psi+ X(\psi) + {1\over 2}\psi\, \div(X) =0.
\end{equation}
This {\em amplitude flow equation} implies the density flow equation and is suggested by the formalism of quantum geodesics when applied to a classical manifold\cite{Beg:geo,LiuMa1,BegMa:cur}.  

Note that if we define the {\em convective derivative along} $X$ as in fluid mechanics by
\begin{equation}
  \frac{D f}{D s} = \dot f+ X(f),\quad   \frac{D Y}{D s}  =  \frac{\extd Y}{\extd s} + \nabla_X Y
\end{equation}
for a function $f$ or another vector field $Y$, then (\ref{velocityflow}),(\ref{densityflow}),(\ref{ampflow}) appear as
\[ {D X\over Ds} =0,\quad   {D \rho\over Ds} =-{\rm div}(X)\rho,\quad  {D \psi\over Ds} =-{1\over 2}{\rm div}(X)\psi.\]
One also has that the metric length of $X$ and its divergence obey\cite{BegMa:cur}
\begin{equation}\label{eq:derNormXsq}
    \frac{D |X|^2}{D s} = 0 ,\quad 
\frac{D \mathrm{div}(X)}{D s} = -(\nabla_\mu X^\nu) (\nabla_\nu X^\mu) - X^\mu X^\nu R_{\mu\nu},  
\end{equation}
in local coordinates, where $R_{\mu\nu}$ is the Ricci tensor. The second of these says that the Ricci tensor controls how a body changes shape (or the stress it experiences) as it freely falls. The first equation says that if the metric length $|X|^2=-1$ for a time-like flow at time $s=0$, then it remains so throughout. This is assumed throughout in \cite{KumMa}. 

The new results in \cite{KumMa} are about demonstrating how this theory plays out around a black-hole. As well as matching to a statistical picture of a large number of geodesics, it is proposed that one natural way to choose the initial $X(0)$ is  by ${\rm div}(X(0))=0$ with prescribed flux on a boundary. Also, by using Kruskal-Szekeres coordinates throughout, there is no problem to follow a Gaussian bump through the horizon and indeed right up to where it impacts the singularity. As a particularly novel feature\cite{KumMa}:

\begin{itemize}\item When two $\rho$ density bumps, collide they typically merge to a bigger bump, but when two $\psi$ amplitude bumps of opposite sign collide they cannot cancel and instead settle into a real dipole with a very different $|\psi|^2$ profile.
\end{itemize}

Hence, the novel idea that $\rho=|\psi|^2$ for an amplitude $\psi$ as a wave function on spacetime (but different from $\varphi$ subject to the Klein-Gordon flow before), if true in some situation, would have a distinctive experimental signature. There is no reason at the moment other than a mathematical one (as the classical limit of a quantum geodesic) for this necessarily to be the case, but it could potentially apply in some context. We are also able to answer what is $s$ at least in this classical case and by extension to the quantum case: it is the time experienced by any particle in the geodesic flow. This is like not one observer but a field of observers which together define the time coordinate $s$. This still takes some getting used to, but should provide a useful new way of thinking about matter. Knowing the flow, one can then relate $s$ to any other local coordinate or laboratory time for a more conventional point of view on the evolution. 
 
\begin{problem}\rm (a) Study classical quantum geodesics in FLRW cosmological models. Also study the role and applications of classical quantum geodesics with $B=C^\infty(N)$, where $N$ could be higher dimensional (these relate to geodesic submanifolds\cite{Beg:geo}). (b) Find an example of a quantum geodesic where $B$ is a QRG such as $\C_q[\R]$, $\C(\N)$ or $\C(\Z)$ with their standard differential calculus. \end{problem}

\section{A fresh look at old problems}\label{sec6}

So far, I have described some recent applications to physics of the QRG programme. Now I look to some of the many challenges that remain more broadly and where I think we might make inroads. 

\subsection{$\kappa$-Minkowski spacetime and curved momentum space} 

The bicrossproduct model spacetime\cite{MaRue}
\[ [x_i,t]=\imath\lambda_p x_i\]
(if we suppose Planck scale corrections) takes its name from the structure of its Poincar\'e quantum group $\C[\R\cross\R^3]\bicross U(so_{1,3})$. Here, the Hopf algebra was introduced by Lukierski et al\cite{Luk}, but by putting it into this form we were able to show that it acts on the above noncommutative spacetime coordinate algebra. One can also use $\kappa=1/\lambda_p$ as the original deformation parameter. The quantum spacetime here is of a class $U(\cg)$, for any Lie algebra $\cg$, viewed since the 1970s as the quantisation of the Kirillov-Kostant Poisson bracket on $\cg^*$. The spin model $U(su_2)$ in \cite{Hoo,FreMa} relevant to 2+1 quantum gravity is also of this form and has quantum double $D(U(su_2))$ as Poincar\'e quantum group. I refer to \cite{Ma:non}  for more details. One of my innovations from that era was to point to a `quantum Fourier transform' 
\[ \CF: C^\infty(G) \to \overline{U(\cg)},\quad \CF(f)=\int_G \extd g f(g) g,\]
where we use the Haar measure and $g=e^{\imath \xi}$ generated by a Lie algebra element $\xi$ in viewed in a completion $\overline{U(\cg)}$. There are ways to make this fully precise (e.g. using von-Neumann algebras), but the take-away for Physics is that we can thereby view $G$ as the classical non-Abelian and hence `curved' momentum space underlying $U(\cg)$ as noncommutative spacetime. One can also refer to this as {\em cogravity} for curvature in momentum space, as dual by some kind of (microlocal) Fourier transform to noncommutative spacetime. Then $U(\cg)$ is thought of as flat spacetime, but in 2+1 quantum gravity with cosmological constant it becomes $U_q(su_2)$ in place of $U(su_2)$, and this is non-cocommutative or `curved' according to the philosophy in \cite{Ma:pla}, and the corresponding momentum space under quantum Fourier transform (which still works) is $\C_q[SU_2]$, i.e. now noncommutative as well as curved, as mentioned in Section~\ref{sec3.4}. We have come a long way using QRG to more general models that are both curved and noncommutative, but the nice class of models above is still of interest. 

One of these reasons to be interested is that the existence of a deformed Poincar\'e quantum group allows one to write down a wave operator as the action of a quadratic Casimir, which for the bicrossproduct model also arises from a QRG as an actual wave operator. This then implied\cite{AmeMa}, under a normal-ordering hypothesis for the correspondence of noncommutative plane waves with observed ones,  a variable speed of propagation for a scalar field, and by assumption also of light. There were, moreover, measurable tests of this using $\gamma$-ray bursts data. 30 years later, there are now enough data points to be able to say that looks promising\cite{Ame0,Ame}. The second of these adds limited data from neutrinos emitted from such events. It raises on the theoretical side:

\begin{itemize}\item What about the actual propagation of light as a $U(1)$ gauge field and of fermions as spinors in the QRG associated to the bicrossproduct model?

\item The Universe at large is not flat, so what happens to the propagation and the analysis in a curved quantum spacetime, for example in a similarly deformed FLRW model?  

\item A difficult part of the analysis in \cite{Ame0,Ame} is to allow for redshift in the data analysis. How to do this while allowing for spacetime to be noncommutative?

\item Can we replace the normal ordering hypothesis by something more geometric and physically justified? 
\end{itemize}

The first item is a hot topic so let me just say that the QRG formulation of fermions goes through Connes axioms of a spectral triple\cite{Con95}, but which in QRG we want to be geometrically realised by an actual (quantum) spinor bundle with connection and some Clifford action data. Meanwhile, the QRG formulation of $U(1)$ gauge theory is well-understood and known already in the case of a trivial bundle for any noncommutative $*$-algebra $A$ equipped with a calculus. See \cite{MaSim2} for some recent work. This geometric side then needs to be tied up with the representation theory of the  Poincar\'e quantum group. An example this last step is the treatment in \cite{MaMc} of a quantum spacetime given by a finite group algebra $\C G$ with $D(G)$  Poincar\'e quantum group. This is both a discrete model in line with 2+1 quantum gravity, and of interest as the Kitaev model in topologically fault-tolerant quantum computing\cite{Kit}. 

On the second item, while one can write down fuzzy FLRW (and fuzzy black-hole) spacetimes where the sphere at a given time and radius is replaced by a fuzzy one\cite{ArgMa3}, such models are not deformations of the above one and have a different flavour. Here I note that the 2+1 bicrossproduct model (interpreted differently at the time with the quantum group viewed as a Heisenberg algebra) was in \cite{Ma:pla} but with the current interpretation was shown in \cite{MaOse} to be twisting-equivalent to the 2+1 quantum gravity model based on $U(su_2)$. Hence the two are cousins and we can transfer back the case of cosmological constant and model quantum spacetime $U_q(su_2)$ to a q-deformation of the bicrossproduct one. Also note that Lukierski et al. \cite{Luk} proceeded by contraction of $U_q(so(3,2))$, so another approach could be to not make this contraction and look for a suitable $q$-deformed model spacetime on which this should act. One can look at this first at the structure of the $2+1$ case\cite{MaSch} to reconcile these approaches.  Solving this issue could then lead into a solution of the 3rd item but would need engaging with a cosmologist to understand the classical level. 

The last item, however, is both the most puzzling and also the one that has to be solved before all of the above can really be said to make contact with physics. As a hint:

\begin{itemize} \item Even if the model and the physics are invariant under a Poincar\'e quantum  group, what does that mean in terms of different frames of reference? The coordinate algebra of the quantum group coacts on the quantum spacetime algebra but all of this is as an algebra: one would need a physical representation as operators of both algebras and equip them with positive linear functionals or states to turn a frame rotation into expectation values and hence actual numbers. A systematic treatment of this issue would not seem impossible, and indeed both algebras have (more or less) reasonable choices of `integrals' as candidates for the ground state.  
\item Quantum geodesics provide a coordinate invariant tool. As first look at these on the bicrossproduct model spacetime was in \cite{LiuMa1}, which should be revisited in a second pass at the calculations.
\end{itemize}
Here, a key result in \cite{LiuMa1} was that if we look at the quantum geodesic flow of a normal ordered Gaussian bump in the spacetime algebra with standard deviation $\sigma$, then (a) expectation values are consistent with speed of light corrections and (b) these corrections are controlled by $\lambda_P/\sigma$.  What the latter means is that there is no such thing as a point particle in the model because any attempt to set $\sigma\to 0$ would imply infinite quantum corrections; we need to keep $\sigma>\lambda_p$. Put another way, quantum gravity corrections at the Planck scale would appear to screen any putative point source and force it to be at least of Planck scale size. Some of this may tie in to other aspects of the bicrossproduct model both going back to my original work\cite{Ma:pla}. One of these is that the underlying deformed action of the Lorentz group on the curved momentum space has an accumulation point which translates as an upper (Planckian) bound on the momentum, making the model `doubly special'. The other is that as part of the construction, the curved momentum space as a group acts back on the Lorentz group as a curved space. The physical meaning of this duality (which was my motivation  for bicrossproducts in the first place in \cite{Ma:pla}) remains to be further understood. This dual model would then provide the physical picture of the coordinate algebra of the Poincar\'e quantum group acting on $U(so_{3,1})$ as a dual spacetime which could be relevant to the first item above. 

Finally, some words about twisting. If we throw away the  Poincar\'e quantum group and just regard the quantum spacetime as an algebra, it is a trivial matter to view it as a 1-sided twist. Here, Drinfeld introduced the notion of conjugating the coproduct as an equivalent quasi-Hopf algebra. A spin off, also called a Drinfeld twist was\cite{Ma} to impose a cocycle condition so that a $*$-Hopf algebra remains one. In my own works from that era, a key result (coming from the equivalence of module categories of the Hopf algebra under a twist) was that there then a 1-sided twist of any $*$-algebra on which the quantum group acts, and indeed on its exterior algebra $(\Omega,\extd)$, see \cite{Ma,Ma:euc, MaOec} and references therein. This geometric side is not part of Drinfeld's theory, however. So, for the case in hand, we can start classically, apply a twist and obtain the same quantum spacetime but now covariant under a different  Poincar\'e quantum group, notably a triangular Hopf algebra obtained from a Drinfeld twist of the classical Poincar\'e algebra. Its modules are now a symmetric monoidal category (i.e. generalising the notion of a super vector spaces). This extends to whatever part of the geometry is invariant such as metrics and connections \cite{BegMa:twi}, which work looked more generally even without the cocycle assumption to give quasi-associate geometries, including the Octonions as such. See \cite[Chapter 9.4]{BegMa} for a recent treatment. 

On the other hand, so long as everything is covariant under the original symmetry, the whole model is not really changed due to the  equivalence of categories. Aschieri, Schenkel and collaborators (in ideas that go back to Julius Wess) have shown\cite{AscSch}, however, how one can still proceed if metrics and connections are not invariant, by including an action on the space of these. Reconciling this with QRG  could be interesting, possibly with a weakening of the axioms of a quantum metric (e.g. with $(\ ,\ )$ not needing to be a left-module map). Also, while doing constructions in a symmetric category is not hard, in the spirit of \cite{Man},  it could  then lead into actual braided geometry in braided categories as arising in the theory of braided groups a.k.a. braided-Hopf algebras\cite{Ma,Ma:alg}. There is also some related work for the geometry of Hopf algebras\cite{AscWeb}, among other works. All of these are interesting directions but one should not loose sight of the fact that such models do not relate to the physics of the bicrossproduct model with its very different Poincar\'e quantum group. On the other hand, from Drinfeld's theory, $U_q(so(3,2))$ is a noncoassociative twist of $U(so(3,2))$ and conceivably this could have a remnant in the contraction process so that the $\kappa$-Poincar\'e quantum group might yet be a twisting in some generalised sense. 

\subsection{Quantum field theory on noncommutative and discrete geometries}

There are plenty of attempts at quantum field theory on noncommutative spacetimes, including the bicrossproduct model one. Deformed Feynman rules are easy enough to write down, e.g. \cite{AmeMa}, and one can also find interesting phenomena such as UV/IR mixing\cite{Gro}. There is a large body of literature here, which I won't attempt to review, but as far as I can see there is still no answer to a fundamental interpretational question: 

\begin{itemize} \item In particle scattering in quantum field theory, we have momenta coming in and momenta going out, with conservation of the total momentum. But if the momentum group is nonAbelian, how do we make sense of the total momentum, as it depends on the order? Equivalently, in a Fock space construction, the order of tensor products matters as the 1-particle Hilbert space has an action of the Poincar\'e quantum group, but the latter is not quasitriangular so that its module category is not even braided. 
\item Such issues do not prevent us from writing down (putative) functional integral versions of the quantum field theory particularly in QRG, where we can write down noncommutative Lagrangians and integrate using a positive linear functional $\int$ on the quantum spacetime. We already  saw this for quantum gravity models in Section~\ref{sec2} and one can similarly look at matter fields and  compute correlation functions. But how does this relate to a Hamiltonian operator quantisation?
\end{itemize}

Regarding the first item, $U_q(so(3,2))$ {\em is} quasitriangular as part of Drinfeld's theory, hence some remnant of this could be visible  in the contraction limit as a generalisation of a braiding in the category of representations. We can also lift everything to the level before we make contractions, which could also answer some of our earlier questions. The second item is much tougher, but worth solving for a decent class of models, such as discrete ones. We have already described results for quantum gravity on a square graph in a functional integral approach. Also known is such an approach for scalar QFT on the lattice line with general QRG metric\cite{Ma:par}, with results including Bekenstein-Hawking radiation for the case of an incoming flat metric at one end of the line and an outgoing flat metric as the other end. To complete the calculation, one has to assume that the relevant plane waves have associated annihilation and creation operators, and for the case in hand there were reasonable choices. However, a general scheme that gives a systematic procedure would be needed to fully justify the steps and to extend the ideas to more general models. Similarly, one can go quite far with quantum gauge theory on a lattice or finite graph\cite{MaSim2} from a functional integral approach, but with the Hamiltonian picture lacking. 

It is worth noting that \cite{MaSim2} also considers finite gauge groups or, more precisely, group algebras, for the gauge symmetry. The distinction is critical for, like in quantum computing, one is replacing elements (here, group elements) by linear combinations of them. On the other hand, if the real world has a quantum group gauge symmetry linked to the quantum spacetime (as in some class of models including 2+1 quantum gravity) then the structure group {\em has} to be a Hopf algebra. Hence this is what we also have to do even for a finite group, for consistency with the full picture. As a result, the symmetries of the theory are much bigger than the case of just the finite group and not its group algebra. Indeed, the theory then decomposes into matrix blocks of noncommutative $U(d_\rho)$-gauge theories\cite{MaSim2} according to the irreducible representations $\rho$ with dimension $d_\rho$ of the finite group. For example, $\C S_3$-gauge theory where $S_3$ is the group of permutations of 3 objects, gives $U(2)\times U(1)\times U(1)$, both for a noncommutative base and indeed even for a classical base manifold. This could give a new take even on the Standard Model. Another remarkable discovery\cite{MaSim2} when we use the group algebra, is the emergence of `gravity-like' modes for the real part of the parallel transport. This is not exactly gravity even in a QRG sense, but suggests the possibility of a unification with gravity in future work. The analysis in the paper is, however, for gauge theory with no restrictions from the differential calculus on the structure group (in algebraic terms, we work with the universal calculus there, but a general calculus on the base). How exactly gauge theory works with general calculi on the structure group is an open problem, even for a tensor product bundle. See the discussion and a partial result in \cite{MaSim2}. 

Next, we note that in the physics literature, there is certain amount of interest in the `Heisenberg double' a.k.a. Weyl algebra of a quantum group $H$. This is a cross product $H\cross H^*$ for a suitable dual, by the left-coregular action (given by evaluation against the coproduct of $H$). However, less well-known it seems is a result in \cite{Ma} that this is isomorphic to ${\rm Lin}(H,H)$ as an algebra by composition of linear maps. So, as before, if one throws away quantum symmetries and other structures then there is little left  to justify one construction over another. In the Lie group case of $H=C^\infty(G)$, the algebra $C^\infty(G)\cross U(\cg)$, in a semi-algebraic version, is the algebra of differential operators $\CD(G)$, and indeed should be thought of a  global version of the Heisenberg algebra. This is a Hopf algebroid, and the same is true for any quantum group as base, as part of an `action Hopf algebroid' construction. Actually doing physics in a systematic manner (as opposed to something ad-hoc) is less clear, but one could in principle repeat generally covariant quantum mechanics\cite{BegMa:qm} as in Section~\ref{sec5} in the case where spacetime is $U(\cg)$, as well as more generally for other quantum groups.

Finally, the Hopf algebroid dual of the algebra of differential operators on a classical manifold is the jet bundle. Applied to the Weyl algebra, one just gets $H \tens H$, the trivial `pair Hopf algebroid', but again this does not take account of the differential calculus. This is also the problem of thinking of the Weyl algebra as the algebra of differential operators - what differentials? In classical geometry, there is a unique translation-invariant calculus on a Lie group and we would quotient the pair Hopf algebroid according to this, but this is not so in the quantum group case. We refer to \cite{HanMa} for details and steps towards the noncommutative jet bundle in this manner. The Hopf algebroid of differential operators $\CD(A)$ in noncommutative geometry is itself unclear for a general algebra $A$ with differential calculus, so we can't just dualise a known theory. In-roads here are in \cite{BHM} building on work of Ghobadi\cite{Gho}.

Meanwhile, a direct approach to the  jet bundle in noncommutative geometry appeared in \cite{MaSim1}. For scalar fields, this is built on $k$-jets
\[ \CJ^k=\oplus_{j=0}^{k} S^j,\quad  S^j=\ker(\wedge_1,\cdots,\wedge_{j-1})\subset (\Omega^1_A)^{\tens_A^j}\]
as an $A$-bimodule (i.e., for sections of the $k$-th jet bundle). Here, the symmetric cotensors $S^j$ were defined as the joint kernel of the adjacent wedge product maps in the j-fold tensor product $\Omega^1\tens_A\cdots\tens_A\Omega^1$. One has a projective limit
\[ \CJ^\infty\to  \cdots \CJ^{k+1}\to \CJ^j\to\cdots \to \CJ^1\to A\]
and $\CJ^1=A\oplus \Omega^1$. There was also a jet prolongation map $j_k:A\to \CJ^k$ built (for higher $k$) with the additional data of a flat torsion free connection. For other matter fields as sections of vector bundles, these appear as $A$-bimodules $E$ and the relevant jet bundle is obtained by $\tens_A E$. Subsequently, \cite{Flo} introduced a very general (and very powerful) `jet-endofunctor' construction which in nice cases reduces to the same carrier space as above, but with a general (recursive) approach to the jet prolongation map that does not need the additional data assumed in \cite{MaSim1}. This data is not a problem in practice, as there are many examples, but how exactly the two approaches play out remains to be seen. 

\begin{figure}
\[ \includegraphics[scale=0.8]{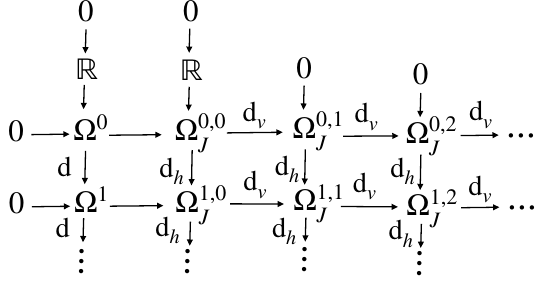}\]
\caption{Variational double complex in the classical case. \label{and}}
\end{figure}

Why we need jet bundles is to define variational calculus. Classically, if $J^\infty$ denotes the total space of the infinite jet bundle, then Anderson and Zuckerman\cite{And,Zuk} proposed to look at $\Omega(J^\infty)$ and require on it the structure of a double complex as in Figure~\ref{and}. So, we need a noncommutative version of $C^\infty(J^\infty)$ and then an exterior algebra on it. As for any vector bundle, functions on the total space what are polynomial in the fibre direction can be viewed as sections of the symmetric tensor products of the dual bundle, so a natural candidate for functions, i.e. in degree 0, is
\[ \Omega^0_J={\rm Sym}_A((\CJ^\infty)^*)={\rm Sym}_A(\CD_\infty)\]
for some appropriate notion of differential operators as dual to $\CJ^\infty$ and some appropriate  notion of symmetrizaiton over $A$. There are candidates for the latter in the setting of \cite{MaSim1} which could be a starting point.  The further information in a double complex is a bigrading, where the total differential is the sum of a `horizontal part' $\extd_h$ inherited from the calculus on $A$ and a `vertical' part $\extd_v$ for differentials in the field space direction. For some convention reasons, we actually extended the field space horizontally in the figure, with $(\Omega,\extd)$ included as a further column on the left. It is fair to say that the problem is somewhat open at this level of generality, bit in my opinion needs to be solved for development of field theory and quantum field theory on a quantum spacetime in a systematic manner. 

Although this is quite abstract and unsolved in general, it can be carried through quite concretely for the case of $A$ the functions on a lattice\cite{MaSim3} as well as presumably (in principle) any other discrete geometry on a graph by the same method. To be really concrete, I will recap the main result of \cite{MaSim3}  in only the simplest case of a lattice line $X=\Z$ with its 2D calculus and basis $\{e^a\}$ of left-invariant 1-forms. We let $I=(a_1,\cdots,a_j)$ be a multindex,  $u_{a_1\cdots a_j}$ a totally symmetric tensor and $e^I=e^{a_1}\cdots e^{a_j}$ in the tensor algebra over $A$. Taking symmetric tensors for the coefficients here is equivalent to keeping $I$ in a standard form $a_1\le a_2\le a_j$. Following \cite{MaSim3}, the total space of the jet bundle has coordinates $i\in\Z$ and the $u_I$, so  
\[ \CJ^\infty=C(\Z)[u_I],\quad j_\infty(\varphi)=\varphi+ (\del_a\varphi)e^a+\cdots = \sum_I (\del_I \varphi) e^I,\]
where we adjoin commuting generators $u^I$ and where the partial derivatives $\{\del_a\}$ for the calculus on $\Z$ are extended to $\del_I$ as the corresponding product. See \cite{MaSim3} for the full explanation, but the idea is quite simply that the jet bundle is a receptacle for the field and all its derivatives to be treated independently, with the jet prolongation map extending a field to all its derivatives. The procedure is similar for a lattice. Some highlights from \cite{MaSim3} are:

\begin{itemize}\item One cannot treat these new variables in the jet bundle classically as we need non-commutation relations, e.g in the lattice line case  $[\extd u_I, u_J]=\sum_{a} u_{aI}u_{aJ}e^a$ (among others).
\item Working in this extended $\Omega_J$, we can define a Lagrangian as form of horizontally top degree and derive Euler-Lagrange equations working in the double complex.
\item In nice cases, as here, one can define exactly on-shell-conserved Noether currents with corresponding Noether charges.
\end{itemize}
This identifies the physically conserved quantities at least  at the level of classical field theory on a lattice. 
For example, on the $\Z$-lattice one can write the standard lattice line equations of motion $(\Delta_\Z+m^2)\varphi=0$, where 
\[ \Delta_\Z=\del_++ \del_-,\quad \del_\pm(\varphi)(i)=\varphi(i\pm 1),\]
in Euler-Lagrange form in $\Omega_J$ and identify the on-shell-conserved energy (thinking of $\Z$ as time, i.e. in 1+0 dimensions) as\cite{MaSim3}
\[ E[\varphi]= -{1\over 2}(\del_+\varphi)(\del_-\varphi) + {m^2\over 2}\varphi^2.\]
In higher dimensions, we also have on-shell-conserved momentum for translation in other lattice directions, and a modified dispersion relation if we make some parallel assumptions to the classical case (this is an issue even in the classical case since the proper  justification comes from quantum field theory not classical field theory). This work in principle leads into the natural way to do quantum field theory on a lattice led by the correct identification of the physical quantities. 

It already seems clear that the result will not be the same as attempts at lattice QFT in the physics literature, which tends to be guided by landing in the continuum limit, whereas in our approach the lattice is an exact discrete noncommutative geometry and we are looking for an exact theory in its own right. This brings us back to why we even need this. Outside of the challenge of lattice QCD etc., my view in these notes is that spacetime could be better modelled as noncommutative/discrete as a way to include some quantum gravity corrections. If so, we then need to redo quantum field theory, including quantum gravity, built on the quantum spacetime. We have already seen some calculable baby models, but to be fully convincing I would propose the following `wish list' of what one would ideally like to see:

\begin{itemize}\item Complete the circle: in a quantum gravity theory on quantum spacetime, are resulting expectation values consistent with the relations of the proposed quantum spacetime? 
\item Use variational calculus to identify the physically relevant Ricci tensor for QRG from variation of the action, and not the other way around as at present.
\item Can we understand quantum geodesics in terms of a quantum variational calculus (generalising the classical notion of a geodesic as extremising the arclength)?
\item A systematic framework for renormalisation for QFT on a quantum spacetime.
\end{itemize}

Regarding the last item, the noncommutatvity parameter or discretisation scale will typically regulate some of the theory and we may not need to renormalise, taking the view that the Planck scale is a natural UV cut-off. However, as we saw in Section~\ref{sec2}, there can still be further divergences and associated field-strength renormalisation needed in principle. Whether we see a generalisation of  the combinatorial Hopf algebras that encode overlapping divergences in the Connes-Kreimer approach for continuum QFT\cite{CK} is unclear. It could be interesting to see what, if anything, is their analogue for a lattice, where renormalisation should normally also be related to coarse-graining. 

\subsection{Outlook} 

While I have focussed on progress as I see it within the QRG programme, there are many other ideas and elements of progress in other approaches to noncommutative geometry, which I have not had the space to mention. For some of these, I refer to a much more extended bibliography in \cite{BegMa}. Even within the physics literature on quantum spacetime, there is, unfortunately too much to individually review but suffice it to say that this has become a vibrant field with many other interesting ideas and speculations. 

Finally, I have in several places mentioned links to quantum computing, such as the Kitaev model\cite{Kit,MaMc,CowMa1,CowMa2}. Looking at the matter from a bigger perspective, the `quantum revolution' here is to replace classicsl states by linear combinations and proceed in a linearised manner. This fits perfectly with the point of view of noncommutative geometry and suggests a much more wide-ranging interaction between these fields. For example:

\begin{itemize}\item Machine learning is about implementation of methods of steepest descent to optimise a benefit function, but if the data is of a quantum nature then we might want to use a quantum differential calculus or even (potentially) quantum geodesics.
\end{itemize}

Another, related, aspect is that computer scientists appear happy to use categorical techniques, in which case an obvious thing to do would be to replace vector spaces by objects of a braided category. One can indeed extend many elements of ZX-calculus to this case, see \cite{Ma:zx} using braid and tangle diagrams to build quantum circuits in which information flows down the page. The difference now is that one has a nontrivial braid operator when two wires cross. The same is useful in QRG to represent a generalised braiding in the construction, e.g. for the tensor product of bimodule connections\cite{BegMa} or for jet bundles\cite{MaSim1}. While such `wiring diagrams' or graphical calculus is routine in quantum computing to provide a nice picture, it was essential back in the early 1990s when it arose in the braided case in my work on braided Hopf algebras \cite{Ma:alg}. For example, early proofs (such as properties of the antipode) were all done this way. A more recent innovation on the categorical side, which I have not had room to say much about, is the notion of a bar monoidal or braided bar category\cite[Chap.~2.4]{BegMa} to properly encode complex conjugation. Suffice it to say that,  looking forward, I see enormous synergies and cross-fertilisation between ideas for physics using noncommutative geometry and the many ideas being explored in quantum computing.  Finally, although I do not yet see a direct link,  it might be hoped that ideas about quantum information might lead back to calculable QRG models for Penrose-Diosi gravitational state reduction as another application in the relatively near future.





\section*{Acknowledgements} This work was supported by Leverhulme Project grant RPG-2024-177. I also want to thank Edwin Beggs, Sam Blitz, Kaushlendra Kumar and Chengcheng Liu for discussions during the respective joint works reported here.

\end{document}